 \documentclass[twocolumn,twocolappendix]{aastex631}

\usepackage{graphicx}	
\usepackage{amsmath}	
\usepackage{amssymb}	
\usepackage{placeins}
\usepackage{mathtools}
\usepackage{cuted}

\makeatletter
\newcommand*{\hyperlinkcite}[1]{\hyper@link{cite}{cite.#1}}
\newcommand{\OmK}{\Omega_\mathrm{K}}
\newcommand{\re}{\operatorname{Re}}

\makeatother

\shorttitle{Convective Overstability in protoplanetary disks}
\shortauthors{M. Lehmann, M.-K. Lin}

\begin{document}

\title{Convective overstability in radially global protoplanetary disks I - Pure gas dynamics}

\author[0000-0002-0496-3539]{Marius Lehmann}
\email{mlehmann@asiaa.sinica.edu.tw}
\affiliation{Institute of Astronomy and Astrophysics, Academia Sinica, Taipei 10617, Taiwan}

\author[0000-0002-8597-4386]{Min-Kai Lin}
\affiliation{Institute of Astronomy and Astrophysics, Academia Sinica, Taipei 10617, Taiwan}
\affiliation{Physics Division, National Center for Theoretical Sciences, Taipei 10617, Taiwan}

\begin{abstract}

Protoplanetary disks are prone to several hydrodynamic instabilities. One candidate, Convective Overstability (COS), can drive radial semi-convection that may influence dust dynamics and planetesimal formation. However, the COS has primarily been studied in local models. This paper investigates the COS near the mid-plane of radially global disk models. 
We first conduct a global linear stability analysis, which shows that linear COS modes exist only radially inward of their Lindblad resonance (LR). The fastest-growing modes have LRs near the inner radial domain boundary with effective radial wavelengths that can be a substantial fraction of the disk radius. 
We then perform axisymmetric global simulations and find that the COS's nonlinear saturation is similar to previous incompressible shearing box simulations. In particular, we observe the onset of persistent zonal and elevator flows for sufficiently steep radial entropy gradients. 
 In full 3D, non-axisymmetric global simulations, we find the  COS produces large-scale, long-lived vortices, which induce outward radial transport of angular momentum via the excitation of spiral density waves.  
 The corresponding $\alpha$-viscosity values of order $10^{-3}$ agree well with those found in previous 3D compressible shearing box simulations. However, in global disks, significant modifications to their radial structure are found, including the formation of pressure bumps. Interestingly, the COS typically generates an outward radial mass transport, i.e. decretion. 
We briefly discuss the possible implications of our results for planetesimal formation and for interpreting dust rings and asymmetries observed in protoplanetary disks. 

\end{abstract}

\keywords{Protoplanetary disks (1300); Hydrodynamics (1963); Astrophysical fluid dynamics (101); Planet formation (1241); Planetesimals (1259)}

\section{Introduction}\label{sec:intro}

The proposed formation of km-sized planetesimals in protoplanetary disks (PPDs) via the core accretion model \citep{safronov1972} faces several obstacles, such as the so-called bouncing and drift barriers, inhibiting efficient coagulation of dust grains (e.g. \citealt{blum2018,drazkowska2022}). In order to overcome these, mechanisms to effectively concentrate dust grains in the mid-plane of PPDs are required. The most popular such mechanism is the streaming instability (SI:  \citealt{youdin2005,youdin2007,johansen2007}), which in its nonlinear saturation can produce dense clumps of dust particles, which can subsequently collapse under the action of selfgravity to form planetesimals \citep{johansen2007,bai2010,simon2016,carrera2021a}. However, for small dust grains the clumping by the SI is strong enough only if the metallicity (or dust abundance) takes at least a few times the solar value (\citealt{li2021}).

In the recent years,  the possible occurrence of purely hydrodynamic instabilities in PPDs has received an increased amount of attention. The main reason is that the magneto-rotational instability (MRI: \citealt{balbus1991}), which had long been considered to be the primary source for mass accretion in PPDs, is likely weak or absent in large parts of PPDs, due to low ionization rates \citep{gammie1996,turner2009,armitage2011,turner2014}. While purely hydrodynamic instabilities, most prominently the vertical shear instability (VSI: \citealt{urpin1998,urpin2003,nelson2013,barker2015,lin2015}), the convective overstability (COS: \citealt{klahr2014,lyra2014,latter2016}) and its nonlinear counterpart --- baroclinic vortex amplification \citep{peterson07a,peterson07b,lesur2010,lyra2011,raettig2013} 
are likely too weak to explain global mass accretion rates in PPDs, they can strongly affect the concentration of dust grains, and can thus be active elements in the planetesimal formation process \citep{lesur2023}

In this paper, we focus on the COS. 
Several studies have investigated this instability in terms of its linear growth phase and its nonlinear saturation, which includes vortex formation \citep{lyra2014,klahr2014,latter2016,teed2021}.
In addition, the concentration of dust in COS-vortices was demonstrated in local shearing box simulations of gas including Lagrangian dust particles \citep{lyra2018,raettig2021}. The results of \citet{raettig2021} suggest that vortices induced by the COS are suitable sites for planetesimal formation, as their simulations revealed dust densities exceeding the Roche density, depending on the disk radius under consideration.

However, the aforementioned studies are based on a fully local shearing box framework, where the radial buoyancy that powers the COS is treated as a constant forcing. This means that the possible effects of the ensuing instability on the disk structure are being ignored.
The exception is the simulations of \citet{klahr2014}, which were conducted in a global cylindrical setup, and more recently the simulations of \citet{klahr2023}, which are in principle fully global. 
However, the former simulations merely probed the linear growth of the instability, and, more importantly, they were 2D-axisymmetric, rather than 3D. The latter simulations were axisymmetric and adopted small radial and vertical extents in favor of resolution.

The goal of this paper is to study the COS in a radially global disk setup, by conducting both linear analyses and nonlinear hydrodynamic simulations (2D axisymmetric and 3D). In contrast to local shearing box simulations, global simulations self-consistently consider the disk's background structure, and possible changes thereof in response to the nonlinear evolution of instabilities and their entailing angular momentum transport. This is expected to be particularly relevant in 3D, where angular momentum can be transport radially due to non-axisymmetric structures. Also the concentration of dust grains should be affected by the global disk evolution.
Furthermore, the most unstable linear COS modes found in fully local calculations actually possess a vanishing radial wavenumber (i.e. an infinite radial wavelength), which somewhat contradicts the local approximation in the first place. Radially global linear calculations presented below indeed confirm that the most unstable modes have a radially global extent, but these modes are not captured by a fully local model.

The paper is organised as follows. In Section \ref{sec:model} we introduce our hydrodynamic model of a PPD. In Section \ref{sec:prelim} we summarize the theoretical background of the COS, which will be useful in interpreting our results, and describe its potential relevance to disk observations. In Section \ref{sec:lin}, we perform a radially global linear analysis of the COS. In Section \ref{sec:hydrosim}, we describe the setup and results of our hydrodynamic simulations. In Section \ref{sec:discussion}, we summarise and discuss caveats of our work and prospects for future work.

\section{Hydrodynamic Model}\label{sec:model}

\subsection{Basic equations}\label{sec:eqn}

We consider a  global hydrodynamic model of a PPD consisting of a compressible, non-isothermal gas, governed by the set of fluid equations
\begin{equation}
\left(\partial_{t} + \boldsymbol{v}\cdot\boldsymbol{\nabla}\right) \, \rho =  - \rho \left( \boldsymbol{\nabla} \cdot \boldsymbol{v} \right), \label{eq:contrhog}
\end{equation}
\vskip -0.6cm
\begin{equation}
 \left(\partial_{t} + \boldsymbol{v}\cdot\boldsymbol{\nabla}\right) \, \boldsymbol{v}    =  - \frac{1}{\rho} \boldsymbol{\nabla} P +  \frac{1}{\rho} \boldsymbol{\nabla} \cdot \boldsymbol{S} - \boldsymbol{\nabla} \Phi_* ,
 \label{eq:contvg}  
\end{equation}
\vskip -0.25cm
\begin{equation}
\left(\partial_{t} + \boldsymbol{v}\cdot\boldsymbol{\nabla}\right) \, P =  - \gamma P \left( \boldsymbol{\nabla} \cdot \boldsymbol{v} \right) -\Lambda. \label{eq:contp}
\end{equation}
In these equations $\rho$, $\boldsymbol{v}$ and $P$ are the gas volume mass density, the three-dimensional velocity, and pressure, respectively, at time $t$. We adopt a non-rotating frame with cylindrical coordinates $(r,\varphi,z)$ and with origin on a central star of mass $M_{*}$ with gravitational potential $\Phi_*$. 
In what follows, we neglect the vertical variation of the potential and set $\Phi_*=-GM_*/r$, where $G$ is the gravitational constant, which implies that we consider a region close to the disk mid-plane such that $|z|\ll H$, where $H$ denotes the pressure scale height defined below. Furthermore, the indirect gravitational term, self-gravity, and magnetic fields are neglected. The remaining symbols in the above equations are explained below. 

We assume an ideal gas equation of state with pressure
\begin{equation}\label{eq:eos}
P=  \frac{\mathcal{R}}{\mu} \rho T,
\end{equation}
where $\mathcal{R}$ is the gas constant and $\mu$ the mean molecular weight.
Following \citet{klahr2014}, we assume gas cooling to occur in the optically thin regime and define the cooling function
\begin{equation}
    \Lambda  = \frac{ \mathcal{R}}{\mu} \frac{\rho \delta T}{t_{c}},
\end{equation}
where $\delta T$ is the deviation of temperature from its equilibrium profile specified below. We subsume viscous heating into this cooling function. 
In what follows we work with the dimensionless constant cooling time
\begin{equation}
\beta = \OmK(r) t_{c}    
\end{equation}
where
\begin{equation}
\OmK = \sqrt{\frac{G M_{*}}{r^3}}
\end{equation}
is the local Keplerian frequency. We fix the adiabatic index $\gamma=1.4$.

Finally, 
\begin{equation}\label{eq:stresstensor}
\boldsymbol{S}= \rho \nu \left[ \boldsymbol{\nabla} \boldsymbol{v} + \left( \boldsymbol{\nabla} \boldsymbol{v} \right) ^{\dagger} - \frac{2}{3} \boldsymbol{I} \boldsymbol{\nabla} \cdot \boldsymbol{v} \right]
\end{equation}
is the viscous stress tensor with the constant kinematic viscosity $\nu$.
The symbol $\dagger$ denotes the conjugate transpose and $\boldsymbol{I}$ stands for the unit tensor. 
In our nonlinear simulations, viscous terms are only included to ensure numerical stability, such that $\nu$ is chosen to be very small. 
 In particular, $\nu$ is much smaller than the typical turbulent viscosities  (defined in Section \ref{sec:numerics}) measured in 3D simulations. Moreover, below, we perform linear calculations, including a kinematic viscosity, to describe the effect of external turbulence on the linear COS.

\subsection{Disk equilibrium}\label{sec:disc_equil}

We focus on the region close to the disk mid-plane $z=0$ and neglect vertical gravity, so the equilibrium configuration does not depend on height $z$ away from the mid-plane. In particular, vertical shear is absent, eliminating the possible occurrence of the VSI.
We assume an axisymmetric disk equilibrium with radial power-law profiles for the gas volume mass density
\begin{equation}\label{eq:rho_equ}
    \rho(r) = \rho_{0} \left(\frac{r}{r_{0}}\right)^{-p}
\end{equation}
and temperature
\begin{equation}\label{eq:T_equ}
    T(r) = T_{0} \left(\frac{r}{r_{0}}\right)^{-q}.
\end{equation}
with constant powerlaw indices $p$ and $q$, and reference density $\rho_{0}=\rho(r_{0})$ and temperature $T_{0} = T(r_0)$ at the reference radius $r_0$.

Although the disk model is unstratified, we can define the characteristic disk thickness or pressure scaleheight of a corresponding vertically stratified model via
\begin{equation}\label{eq:scaleheight}
H = c_s/\OmK, 
\end{equation}
where 
\begin{equation}\label{eq:soundspeed}
c_{s}=\sqrt{\mathcal{R}T/\mu}    
\end{equation}
is the sound speed.

The equilibrium velocity of the gas is given by 
\begin{align}\label{eq:v_eq}
\boldsymbol{v}(r)  & = r \Omega(r)   \boldsymbol{e}_{\varphi},
\end{align}
with the orbital frequency 
\begin{equation}\label{eq:omega}
\Omega(r)= \OmK(r)\left[1 - 2\eta(r)\right]^{1/2},
\end{equation}
where we defined the dimensionless radial pressure gradient \citep{youdin2005}
\begin{equation}\label{eq:eta}
\eta(r) \equiv -\frac{1}{2 \rho (\OmK r )^2} \frac{\partial P}{\partial \ln r} = \frac{h^2}{2}\left(p+q\right)
\end{equation}
with the disk aspect ratio $h=H/r$. We also define $\Omega_{0} = \Omega(r_0)$ and $H_0=H(r_0)$.
Furthermore, the squared radial buoyancy frequency is given by
\begin{equation}\label{eq:nr2} 
     N^2  \equiv -\frac{1}{\gamma\rho}\frac{\partial P}{\partial r}\frac{\partial S}{\partial r} =  -\frac{1}{\gamma} h^2 \Omega^2 \left(p+q\right)\left(q+\left[1-\gamma\right]p\right)
\end{equation}
with the dimensionless entropy
\begin{align}\label{eq:ent}
    S \equiv \ln{\frac{P}{\rho^\gamma}}.
\end{align}
The second equality in Equations (\ref{eq:eta})-(\ref{eq:nr2}) assumes the radial power-law profiles defined above.

\section{Preliminaries on the COS}\label{sec:prelim}

\subsection{Theoretical aspects}

The COS comprises an oscillatory destabilization of inertial waves of typical frequency $\sim \Omega$, induced by a convectively unstable radial entropy gradient, where epicyclic oscillations are amplified on account of a radial buoyancy force which results from the cooling of gas occurring on a typical time scale $t_{c}\sim 1/\Omega$ (see \citealt{latter2016} for an illustrative explanation and \citealt{teed2021} for more details).
Under typical conditions where the radial pressure gradient $\partial_r  P<0$, the COS requires a negative radial entropy gradient, $\partial_r S <0$, since then the squared radial buoyancy frequency (\ref{eq:nr2}),
which is the central quantity for the linear COS, is negative.
 The linear instability mechanism behind the COS was first studied by \citet{klahr2014}, \citet{lyra2014}, and \citet{latter2016}. 

The nonlinear saturation of the COS was first explored in compressible 3D shearing box simulations by \citet{lyra2014}. These simulations revealed the formation of large scale vortices and an entailing radial angular momentum transport characterized by $\alpha$-viscous values on the order $\sim 10^{-3}.$
The nonlinear saturation of the axisymmetric COS was studied in detail by 
 \citet{teed2021} (\hyperlinkcite{teed2021}{TL21} henceforth) employing local incompressible shearing box simulations. These authors found that two dimensionless quantities largely determine the nonlinear outcome of the instability. These are the Reynolds number 
\begin{equation}\label{eq:reynolds}
    Re=\frac{L^2 \kappa}{\nu} \equiv \frac{1}{\alpha(r)}, 
\end{equation}
measuring the strength of viscous diffusion, 
and the (pseudo-)Richardson number
\begin{equation}\label{eq:richardson}
    R =  \frac{N^2}{\kappa^2}\equiv h^2(r) \frac{\left(p+q\right)\left(q+\left[1-\gamma\right]p\right)}{\gamma} , 
\end{equation}
measuring the strength of the destabilizing radial entropy stratification against the stabilizing effect of the radial angular momentum gradient. 
Here, $L$ is a characteristic length scale of the flow and $\kappa$ denotes the epicyclic frequency. Since our disk model is (as in \hyperlinkcite{teed2021}{TL21}) vertically unstratified, this length scale should be the vertical extent of the simulation region. Our simulations below adopt a vertical domain $L_z=0.5 H$ (see \S \ref{sec:hydrosim}), but for simplicity we set $L\sim H$, and we use $\kappa\approx \OmK$. Assuming this and considering the power-law disks adopted above gives the second equalities in Eqs. (\ref{eq:reynolds})-(\ref{eq:richardson}). Note that in contrast to \hyperlinkcite{teed2021}{TL21}, our model is radially global, such that $Re$ and $R$ are functions of disk radius. Specifically, $R$ ($Re$) increases (decreases) with decreasing radius, implying a more vigorous COS at smaller radii.

\hyperlinkcite{teed2021}{TL21} found that, depending on the values of  $R$ and $Re$, the COS saturates either into a weakly nonlinear state characterised by ordered nonlinear waves, a state characterised by wave turbulence, or a state supporting intermittent or persistent zonal flows and elevator flows (see \S \ref{sec:sim_2d}). They found that $R$ needs to exceed a critical value, roughly delineated by the relation\footnote{The factor 0.1 instead of 0.05 (as in \hyperlinkcite{teed2021}{TL21}) arises since our scaling length is $H=2 Lz$, rather than $L_{z}$ .}
\begin{equation}\label{eq:ncrit}
R_{\text{crit}} =  0.1 \left(Re/10^4\right)^{-1/2}
\end{equation}
in order for zonal flows to be formed in the saturated state. The point is that the COS is only expected to play a role in dynamical process in PPDs if zonal flows are formed, which for instance can concentrate dust grains. Furthermore, in 3D simulations zonal flows are expected to wrap up into vortices, which can induce radial angular momentum transport, and, which are likewise able to concentrate dust \citep{raettig2021}.

\subsection{Occurrence in PPDs}

\begin{figure}
\centering 
	\includegraphics[width=0.5 \textwidth]{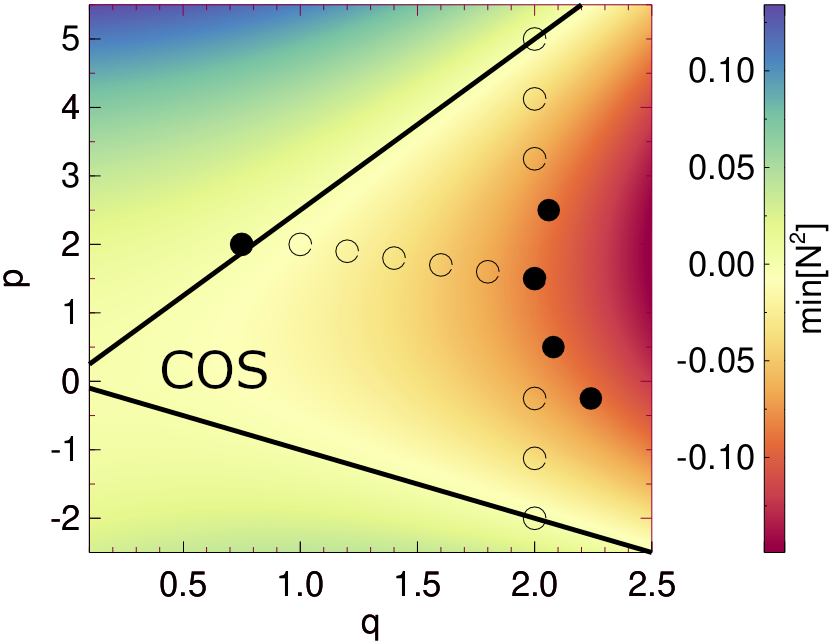}
    \caption{Contour plot showing the minimal value of $N^2/\OmK(r_0)^2$ within a radial region $r_0\pm 2H_0$ in a power law disk with density and temperature slopes $p$ and $q$, respectively. The diagonal lines delineate the regions where $\text{min}\left[N^2\right]$ takes negative or positive values, such that the COS is in principle expected to occur in the parameter-space labeled "COS". Open (solid) circles represent values of $p$ and $q$ used in our 2D-simulations (2D as well as 3D simulations) below.}
    \label{fig:cos_crit}
\end{figure}

The requirement of $N^2<0$ for the COS to operate in the disk mid-plane typically translates to  a rather steep temperature gradient and a flat density profile (see Figure \ref{fig:cos_crit}), which would be inconsistent with global disk profiles inferred from early mm and sub-mm continuum observations of PPDs (e.g. \citealt{andrews2009,isella2009}). However, flat density profiles could be realized, for instance, near the edges of dead zones, planet-carved gaps, or pressure bumps in general. As such, COS-active regions would be expected to have typical radial widths $\Delta r \sim H$. On the other hand, more recent ALMA imaging of HD 163296 \citep{mathews2013} and several disks in Lupus \citep{tazzari2017} revealed very flat surface mass density profiles at disk radii around tens of AU, which should in principle increase the likelihood for the occurrence of the COS.

Whether the COS can operate also strongly depends on the gas' cooling timescale and is thus tied to the distribution of small dust grains tightly coupled to the gas 
\citep{malygin2017,barranco2018,pfeil2019,fukuhara2021}. The above studies attempted to map out the gas cooling time scale across the radial and vertical extent of a PPD. However, such calculations require multiple simplifications, such that the computed cooling times are subject to large uncertainties.  Nevertheless, these estimates suggest that one or more purely hydrodynamic instabilities are likely active in planet-forming regions of the disk. In particular, the results of \citet{pfeil2019} suggest that the COS may occur in the potential planet-forming region at $r=1-10 \text{AU}$ if the structural requirement $N^2<0$ is fulfilled. 

\section{Radially global linear analysis}\label{sec:lin}

\subsection{Linearised Equations}\label{sec:lin_eq}
We now assume the disk equilibrium described in \S \ref{sec:disc_equil}
and investigate the stability of axisymmetric radially global, vertically local perturbations
of the form
\begin{equation}
    \delta \rho \to \delta \rho(r) \exp{\left(i k_{z} z - i \omega t\right)},
\end{equation}
and similar definitions for the perturbations $\delta v_{r}$, $\delta v_{\varphi}$, $\delta v_{z}$ and $\delta P$. We define the complex mode frequency $\omega=\omega_{R} + i \omega_{I}$, such that $\omega_{R}$ and $\omega_{I}$ denote the real-valued mode's frequency and growth rate, respectively. Furthermore, $k_{z}$ is the real-valued vertical wavenumber and we take $k_z>0$ without loss of generality. 
We also define
\begin{align}
    W & = \frac{\delta P}{\rho},\\
    Q & = c_{s}^2 \frac{\delta \rho}{\rho},
\end{align}
for convenience.

The linearised cylindrical equations then read
\begin{align}\label{eq:linQ}
    i \omega \frac{Q}{c_{s}^2}  = & \,\partial_{r} \ln \rho \delta v_{r} + i k_{z} \delta v_{z} + \partial_{r} \delta v_{r} + \frac{\delta v_{r}}{r} ,\\ 
\label{eq:linvr}
   i \omega \delta v_{r} = & -2 \Omega \delta v_{\varphi} - \frac{\partial_{r} P}{\rho c_{s}^2} Q + \partial_{r} W + \partial_{r} \ln \rho W,\\
\label{eq:linvphi}
   i \omega \delta v_{\varphi} = & \left(2\Omega + r \partial_{r} \Omega \right) \delta v_{r},\\ 
\label{eq:linvz}
    i \omega \delta v_{z} = & \,i k_{z} W,\\
\label{eq:linW}
\begin{split}
    i \omega W = & \frac{\partial_{r} P}{\rho}\delta v_{r} + \gamma c_{s}^2\left( i k_{z} \delta v_{z} + \partial_{r} \delta v_{r} 
     + \frac{\delta v_{r}}{r}\right) \\
     \quad & + \frac{1}{t_{c}}\left(W-Q\right).
    \end{split}
\end{align}

Note that the driving force of the COS is the radial buoyancy force, encapsulated in the term $\propto \partial_{r} P$ in the radial momentum equation.

\subsection{WKB dispersion relation}

We can easily derive the WKB dispersion relation corresponding to Eqs. (\ref{eq:linQ})---(\ref{eq:linW}) by ignoring all radial variations of the equilibrium disk structure, as well as curvature terms. Furthermore, we introduce the complex radial WKB wavenumber $k_r(r)$, assuming that $\partial_{r}\to i k_r(r)$.
We then obtain the fifth order dispersion relation
\begin{equation}\label{eq:wkbdisp}
    -i t_{c} \omega^3 + \omega^2 -c_{s}^2 \left( k_{z}^2 + \frac{\omega^2 k_{r}^2}{\omega^2 -\Omega^2} \right) \left(1- i \gamma \omega t_{c} \right) =0.
\end{equation}
In the isothermal and adiabatic limits $t_{c}\to 0$ and $t_{c}\to \infty$ we find
the fourth order dispersion relation
\begin{equation}\label{eq:wkb_isotherm}
    \omega^2 -c_{s}^2\left(k_{z}^2 + \frac{\omega^2 k_{r}^2}{\omega^2-\Omega^2}\right) =0.
\end{equation}
and
\begin{equation}\label{eq:wkb_adiabatic}
   \omega^2 -\gamma c_{s}^2\left(k_z^2 + \frac{\omega^2 k_r^2}{\omega^2-\Omega^2}\right) =0,
\end{equation}
respectively. Note that (\ref{eq:wkbdisp}) should not yield any growing solutions, which can easily be shown in the isothermal and adiabatic limits, since in those cases the dispersion relation yields a simple quadratic equation in $\omega^2$.
Furthermore, in the limit of high mode frequencies $|\omega|\gg \Omega$ we isolate acoustic waves
\begin{equation}
        \omega^2  - \gamma c_{s}^2 k^2=0
\end{equation}
in the adiabatic limit, where $k=\sqrt{k_r^2+k_z^2}$. The isothermal result can then be obtained by setting $\gamma$ to unity. 

On the other hand, in the incompressible limit $c_{s}\to \infty$ we find
\begin{equation}
    \left(k_{z}^2 + \frac{\omega^2 k_{r}^2}{\omega^2-\Omega^2}\right)\left(1-i \gamma \omega t_{c} \right)=0.
\end{equation}
The solution of the latter is the cooling mode
\begin{equation}
    \omega_{\text{COOL}} = -\frac{i}{\gamma t_{c}},
\end{equation}
decaying on cooling time scale $t_{c}$,
as well as the pair of inertial waves
\begin{equation}\label{eq:inertial_wave}
   \omega_{\text{IW}}= \pm \frac{k_{z}}{k}\Omega.
\end{equation}

Since the axisymmetric COS is captured within the incompressible Boussinesq approximation (\citealt{latter2016,lp2017};~\hyperlinkcite{teed2021}{TL21};~\citealt{lehmann2023}), we expect that short wavelength COS modes in our radially global model essentially are inertial waves, described locally by Eq. (\ref{eq:inertial_wave}). This will indeed be confirmed below. 
The Boussinesq approximation can be obtained from Eqs. (\ref{eq:contrhog})---(\ref{eq:contp}) by
assuming  that the length scale of the phenomena under investigation is much shorter than the radial and vertical disk stratification lengths, that the flow is strongly subsonic ($|\delta \boldsymbol{v}| \ll c_{s0}$), and that the fractional perturbations of density $\delta \rho/\rho\ll 1$ and pressure $\delta P/P \ll 1$, but with $\delta \rho/\rho \gg \delta P/P $. The latter assumption is necessary to retain the effect of the background entropy gradient in the model equations (see \citealt{lp2017} for details).

\subsection{Numerical methods}\label{sec:lin_num}
When supplemented with two boundary conditions accounting for the two first order derivatives in $r$, such as
\begin{equation}
    \delta v_{z}(r_{\text{min}}) = \delta v_{z}(r_{\text{max}})=0, 
\end{equation}
 Eqs. (\ref{eq:linQ})---(\ref{eq:linW}) constitute a linear eigenvalue problem of the form
\begin{equation}\label{eq:lineig}
    \boldsymbol{\mathcal{L}} \boldsymbol{b} = i \omega \boldsymbol{b}
\end{equation}
with the linear differential operator $\boldsymbol{\mathcal{L}}$, eigenvector $\boldsymbol{b}=\left\{Q,  \delta v_r, \delta v_{\varphi}, \delta v_{z}, W \right\}^T$ 
and complex eigenvalue $i\omega$.
We use the spectral code Dedalus \citep{burns2020} to solve the linear eigenvalue problem (\ref{eq:lineig}) on a Chebyshev grid using a collocation point method (see, for instance, \citealt{lin2021} for a similar treatment of a vertically stratified disk model).

Complementary to this, we verify the results obtained with the spectral code using an alternative algorithm based on the shooting method.
To this end, we first recast the above equations as two first-order ordinary differential equations in $r$:
\begin{equation}\label{eq:ode_vr}
\begin{split}
    \partial_{r} \delta v_{r}  = & - \left( \frac{1}{r} + \frac{i \,\partial_{r} \ln\rho}{i + \gamma \omega t_{c}} +\frac{\partial_{r} P \omega t_{c}}{P\left(i+\gamma \omega t_c \right)}\right)\delta v_r \\
    \quad & + \left(-\frac{i k_{z}^2}{\omega} + \frac{\omega \rho \left(-1 + i \omega t_c \right)}{P\left(i + \gamma \omega t_c \right)}\right) W,
\end{split}
\end{equation}
\begin{equation}\label{eq:ode_W}
\begin{split}
    \partial_{r}  W  = & i \bigg(\omega -\frac{2 \Omega\left(2\Omega + r \partial_{r}\Omega \right)}{\omega} + \\
    \quad & \partial_{r} \ln P \frac{ \left(\partial_{r} P - \gamma P \partial_{r} \ln \rho \right) t_{c}}{ \rho \left(i + \gamma \omega t_{c} \right)}\bigg) \delta v_{r} \\
\quad & + \left(\partial_{r} \ln P \frac{ \left(i + \omega t_{c} \right) }{\left(i + \gamma \omega t_{c} \right)}  -\partial_{r} \ln\rho \right)W,
\end{split}
\end{equation}
which are then integrated using a fourth or fifth-order Runge-Kutta scheme with pre-specified outer (we shoot from the outer to the inner domain boundary) boundary values for $\delta v_{r}$ and $W$. The solution with the correct value for $\omega$ is then obtained via a Newton-Raphson iteration algorithm to match the corresponding inner two boundary values for one of the fields (we use $\delta v_{z}$), obtained first with the spectral method. We find that both methods generally are in excellent agreement.
However, in certain cases modes possess very short radial wavelengths ($k_{r} H_{0}\gtrsim 100$), such that the spectral method becomes prohibitively slow to obtain fully resolved radial profiles of the eigenfunctions.
Nevertheless, the method can still find the correct eigenvalue $\omega$ corresponding to such solutions, which we verified using example calculations with subsequently increased radial resolution. For such cases, we present radial profiles obtained with the shooting method, which can be computed efficiently even for a large number of radial grid points $n_{r} \gtrsim 10^4$.

\subsection{Results}

\subsubsection{Smooth disk region}\label{sec:lin_res_fid}

\begin{figure}
\centering 
	\includegraphics[width=0.5 \textwidth]{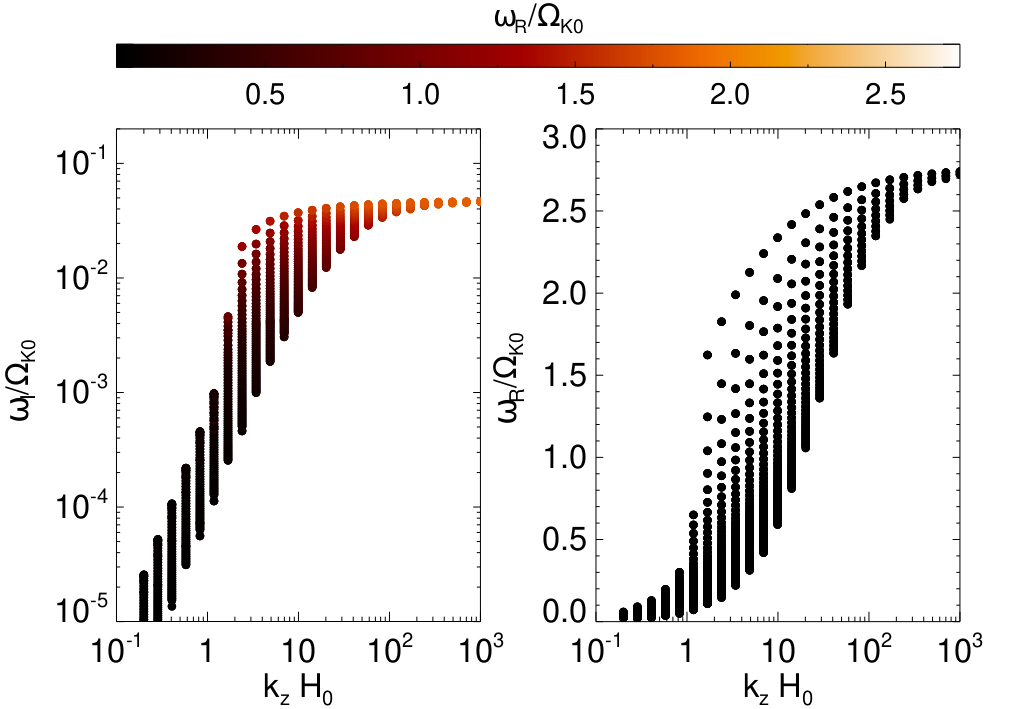}
    \caption{Growth rates  $\omega_I$ and frequencies $\omega_R$ as functions of vertical wavenumber of linear, radially global COS modes in the fiducial disk ($p=1.5$, $q=2$, $5\leq r/H_{0} \leq 15$). }
    \label{fig:disprel_fid}
\end{figure}

\begin{figure*}
\centering 
	\includegraphics[width=\textwidth]{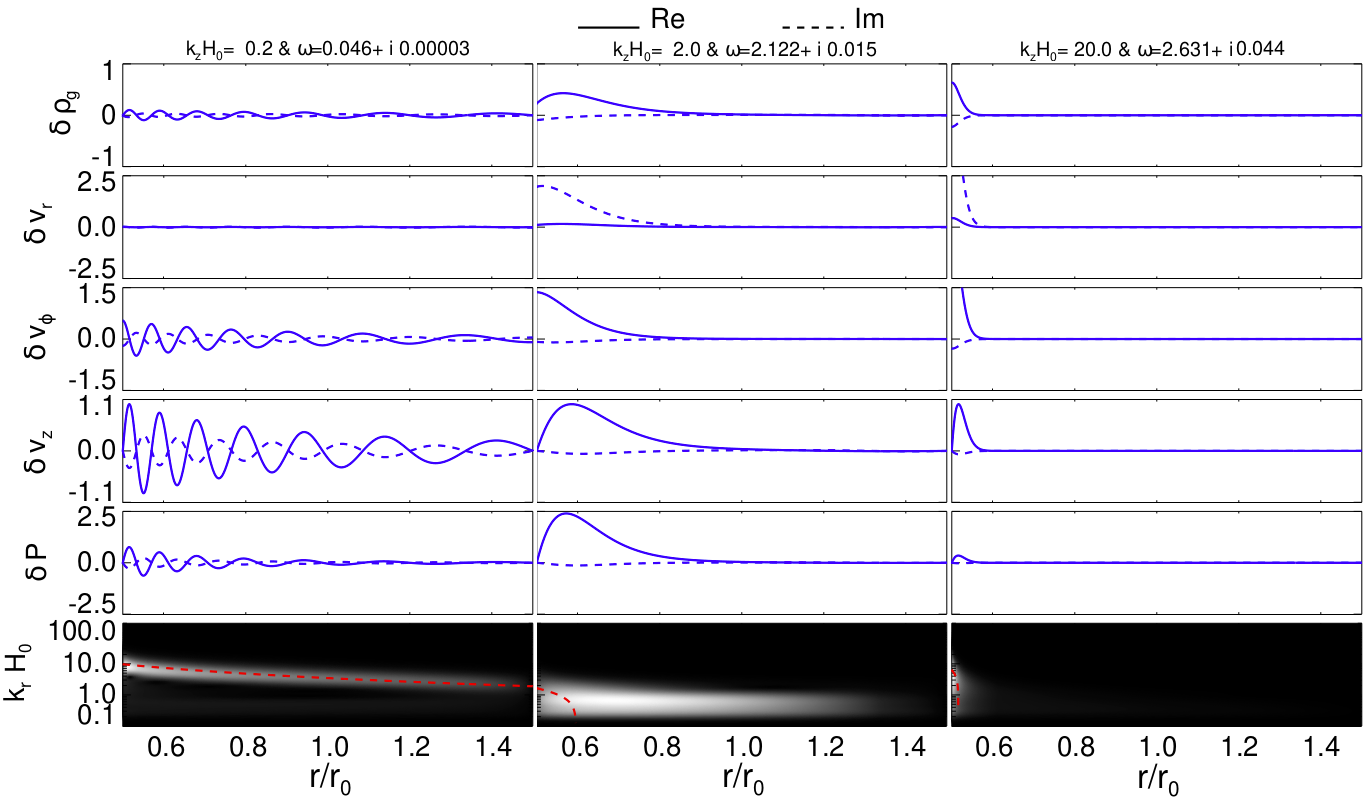}
    \caption{Radial eigenfunctions corresponding to the fastest growing COS modes in the fiducial disk with $p=1.5$ and $q=2$ for increasing vertical wavenumbers $k_{z} H_{0}=0.2,2,20$, as obtained with the spectral method. Notably, the eigenfunctions become increasingly concentrated toward the inner domain boundary with increasing $k_{z}$.  The bottom panel shows the wavelet power corresponding to the radial profile of $\re\left[\delta v_{z}\right]$. The red dashed curves in the wavelet power spectra correspond to $k_{r}$ obtained from (\ref{eq:inertial_wave}), with accordingly specified $k_{z}$ and using $\omega_{\text{IW}}=\omega_{R}$. For the modes with $k_{z}H_{0}=2,20$ the Lindblad resonance defined via (\ref{eq:lindblad}) is located within the computational domain, where the red dashed curves drop to zero.}
    \label{fig:eigen_fid}
\end{figure*}

Figure \ref{fig:disprel_fid} shows frequencies $\omega_{R}$ and growth rates $\omega_{I}$ of linear COS modes for vertical wavenumbers $0.2 \leq k_{z} H_{0} \leq 1,000$, obtained with the spectral code. Here we consider a radial domain $5\leq r/H_{0} \leq 15$. As outlined in \S \ref{sec:intro}, typical regions that are unstable to the COS are expected to be significantly smaller. However, in our simulations below we will adopt the same radial domain, spanning $10 H_{0}$. The reason is that we wish to avoid spurious influences from the radial boundary conditions on the long term nonlinear evolution in our simulations. Moreover, the linear analysis of a smooth disk region does not depend qualitatively on the radial domain size (cf. \S \ref{sec:loc_glob}).

For each value of $k_{z}$ we generally find a band of modes within a range of oscillation frequencies and growth rates, with the average frequencies and growth rates increasing with increasing $k_z$. Moreover, larger growth rates generally correspond to larger oscillation frequencies.
Overall, oscillation frequencies ranging from small values $\omega_{R}\sim 10^{-2}\Omega_{0}$ to roughly the largest value of the Keplerian frequency within the considered domain $\omega_{R}\approx 2.7\Omega_{0}$ (i.e. typical frequencies of local inertial waves with $k_z \lesssim k$) can occur. Note that for smaller $k_z$ the lowest occurring frequency is set by the radial grid resolution of the spectral solver.
Furthermore, modes with lower frequency generally extend further out through the domain, whereas with increasing frequency (and hence increasing growth rate) modes  tend to be increasingly concentrated toward the inner domain boundary. This is illustrated in Figure \ref{fig:eigen_fid}, which shows the radial profiles of the various hydrodynamic quantities involved in the problem. 

The behavior of the modes just described can be better understood if we consider the WKB dispersion relation in the isothermal or adiabatic limit (\ref{eq:wkb_isotherm})-(\ref{eq:wkb_adiabatic}). The latter can be rearranged to yield
\begin{equation}\label{eq:wkb_resonance}
 k_r^2 = \frac{1}{\gamma c_s^2}\left(\omega^2-\Omega^2\right)\left(1-\frac{\gamma c_s^2 k_z^2}{\omega^2}\right).
\end{equation}
This equation is analogous to Eq. (60) of \citet{lubow1993} and Eq.
(12) of \citet{svanberg2022}, the latter applying to vertically stratified, radially local disk models.

According to Eq. (\ref{eq:wkb_resonance}) neutral\footnote{with real-valued $k_{r}$ and $\omega$} waves can only propagate in regions where the two conditions $\omega>(<)\Omega$ \emph{and} $\omega>(<) \sqrt{\gamma} c_{s} k_{z}$ are simultaneously fulfilled. The first (second) equality applies to acoustic (inertial) waves.
Since our linear analysis above reveals no modes with $\omega>\Omega$, we conclude that COS modes only exist radially inwards from their Lindblad resonances (LRs), whose radii $r_{\text{LR}}$ are given by the condition 
\begin{equation}\label{eq:lindblad}
\omega=\Omega(r_{\text{LR}}).
\end{equation}
This appears plausible, since in the Boussinesq approximation the linear COS is known to excite inertial waves \citep{klahr2014,lyra2014,latter2016}, and presumably there exist no linear (global) modes in our disk model which connect both inertial and acoustic waves.

Nevertheless, modes that can appear in our simulations are restricted to vertical wavenumbers $k_{z} H_{0}\gtrsim 10$, such that the modes in our simulations can only fulfill the condition $\omega<\Omega$ and waves can only propagate \emph{inside} of their Lindblad resonance radius. Note that generally, our vertically unstratified analysis is only meaningful for modes with $k_{z}H_{0}>1$.

The above implies that lower frequency modes have their LR at larger radii, such that these modes possess smaller radial wavelengths within the computational domain. Consequently, their growth rates are smaller, as their coupling to the background entropy gradient becomes less efficient.

In addition, we compute a wavelet transform\footnote{We use the IDL 'wavelet' procedure with a mother wavelet of type 'Paul' \citep{torrence1998}.}  
of the vertical velocity $\delta v_{z}$, which enables us to estimate the radial wavenumber $k_{r}(r)$ of the eigenmodes.
The bottom panel in Figure \ref{fig:eigen_fid} shows the wavelet power spectrum as a function of radius and radial wavenumber $k_{r} H_{0}$. 
The mode with $k_{z} H_{0}=0.2$ shows a radial change of its radial wavenumber that is characteristic of global waves, such as spiral density waves in planetary rings \citep{shu1984}. This radial change results from the radially varying $\OmK$.
Note that the eigenfunctions for the cases $k_{z} H_{0}=2$ and $k_{z}H_{0}=20$ are actually not wavelike, i.e. they do not undergo at least one full oscillation in radial direction. Therefore, the wavelet transforms for these cases cannot be expected to yield a clear radial wavenumber.
To improve the results we re-scaled the wavelet power such that its minimal value is set to zero (by manual subtraction) for these cases, as this procedure eliminates artificial residuals in the wavelet power spectrum.
The red dashed curve corresponds to the radial wavelength of an inertial wave (\ref{eq:inertial_wave}) with given vertical wavenumber $k_{z}$ and frequency $\omega_{R}$.

Finally, we note that we checked the robustness of the computed modes by using a number of different boundary conditions. Also we verified the existence of some (including the fastest growing modes) with the shooting method. \\ \\

\subsubsection{Connection to local COS modes}\label{sec:loc_glob}

\begin{figure}
\centering 
	\includegraphics[width=0.5\textwidth]{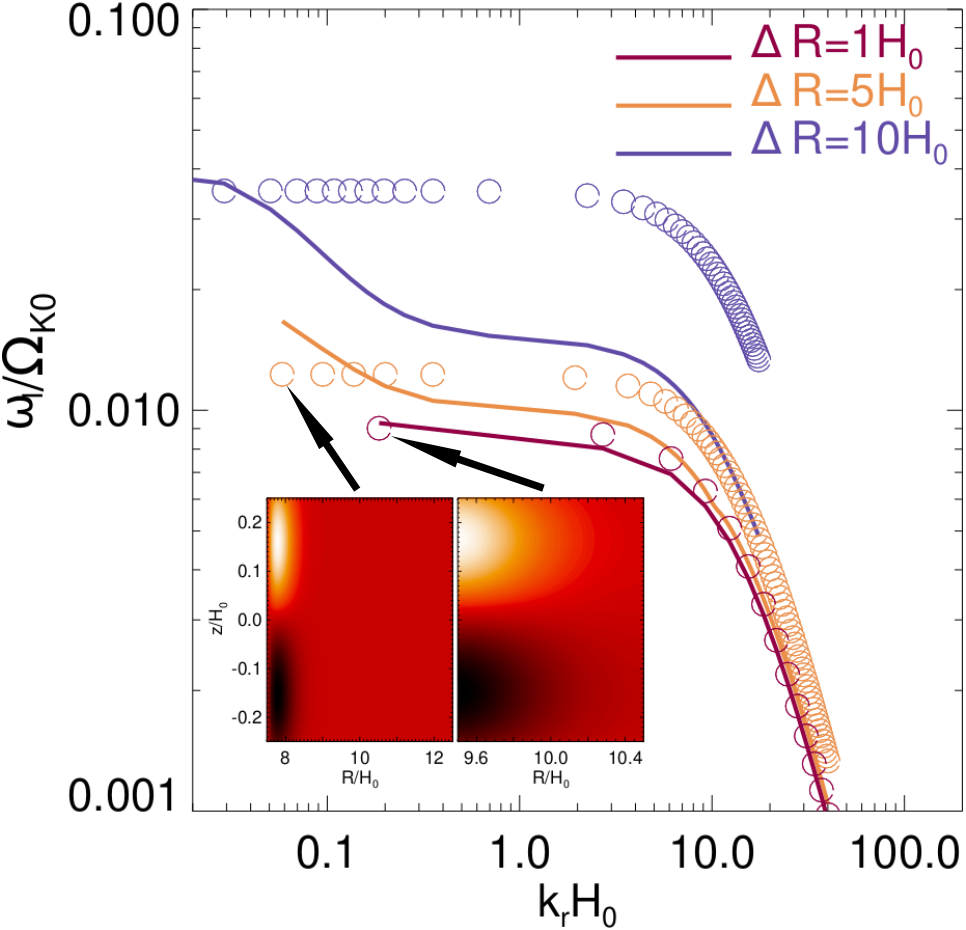}
    \caption{Linear growth rates of radially global COS modes (solid curves) as function of radial wavenumber $k_{r}$ for fixed $k_{z}H_{0}=10$ and for different radial sizes of the computational domain $\Delta r$ are compared with fully local growth rates (circles), computed via Eq. (\ref{eq:gr_loc}). The radial wavenumbers $k_{r}$ are obtained from (\ref{eq:wkbdisp}) and correspond to radial averages. See the text for details. The insert figures illustrate contours of the vertical velocity perturbation of the fastest growing modes in the cases $\Delta r=5 H_{0}$ and $\Delta r=H_{0}$.}
    \label{fig:glob_loc}
\end{figure}

We can attempt to compare the linear growth rates of the global COS modes derived in the previous section with those of fully local modes.
To this end, we consider the expression for the growth rates of local modes with wavenumber $\boldsymbol{k}=k_{r}\boldsymbol{e}_{r} + k_{z} \boldsymbol{e}_{z}$ \citep{klahr2014,lyra2014,lehmann2023}: 
\begin{equation}\label{eq:gr_loc}
    \omega_{I,\text{loc}} = -\frac{\beta \mu^2 N^2 \Omega}{2\left(\Omega^2+ \beta^2 \mu^2\left[\Omega^2+N^2\right]\right)}
\end{equation}
where $\mu = k_{z}/\sqrt{k_{z}^2 + k_{r}^2}$.
This expression has been derived using the incompressible Boussinesq approximation \citep{lp2017}, which yields almost identical results to a local compressible calculation (see Appendix \ref{app:compress}). 
We expect that the global COS modes derived in the previous section increasingly resemble local COS modes as we decrease the radial domain size, since then the variation of background quantities becomes smaller.
This is illustrated in Figure \ref{fig:glob_loc}, where we present growth rates of global modes with $k_{z} H_{0}=10$ for three different domain sizes $\Delta r= H_{0}$, $\Delta r= 5 H_{0}$  and $\Delta r =10 H_{0}$, as a function of radial wavenumber $k_{r}$, represented by the solid lines.
These are compared with local modes following from (\ref{eq:gr_loc}), represented by open circles.
The values of $k_{r}$ for each mode are obtained using the WKB dispersion relation (\ref{eq:wkbdisp}), where we assume values for $c_{s}^2, t_{c}$ and $\Omega$ corresponding to radial averages over the entire domain. Similarly, radial averages of $N^2$ are used in (\ref{eq:gr_loc}). Note that generally the WKB wavenumber $k_{r}$ is complex, and the values in Figure \ref{fig:glob_loc} are the corresponding real parts. 

Thus, for domain sizes $\Delta r \lesssim H_{0}$, short wavelength modes with $k_{r}H_{0}\gg 1$ resemble local COS modes, as indicated by the good match in the corresponding growth rates. However, the fastest growing modes are still global, as they do not extent throughout the entire domain unlike local modes, but are concentrated toward the inner domain boundary. This is illustrated in the insert figure, which shows the corresponding vertical velocity field for $\Delta r=H_{0}$ and $\Delta r =5 H_{0}$. 
\begin{figure*}
\centering 
	\includegraphics[width=\textwidth]{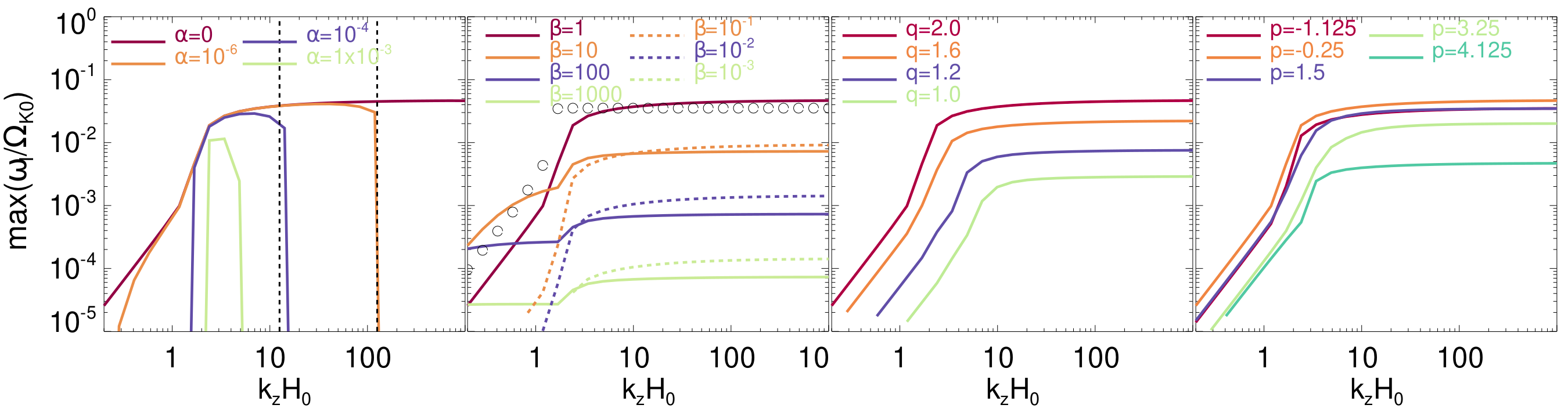}
    \caption{Maximal linear growth rates of radially global COS modes in a smooth disk. Compared is the effect of a varying $\alpha$-viscosity (left panel), cooling time (second panel), temperature slope (third panel) and density slope (rightmost panel). If not indicated otherwise parameters correspond to the fiducial disk ($q=2$, $p=1.5$, $\beta=1$, $\alpha=0$).  Open circles in the second panel from the left correspond to growth rates obtained from a fully local model.}
    \label{fig:eigen_var}
\end{figure*}
The reason is that the fastest growing modes' LR lies within the computational domain, implying that their frequencies $\omega_{R}\gtrsim \Omega_{0}$.
Again, we checked these modes' robustness by using several different boundary conditions.
With increasing domain size the radial variation of the background quantities (particularly $\Omega_{K}$) results in an increasing discrepancy between local and global growth rates, as $k_{r}$ exhibits an increasing variation throughout the domain. Obtaining local growth rates becomes less meaningful for larger domain sizes.

\subsubsection{Dependence on disk parameters}

\begin{figure*}
\centering 
	\includegraphics[width= \textwidth]{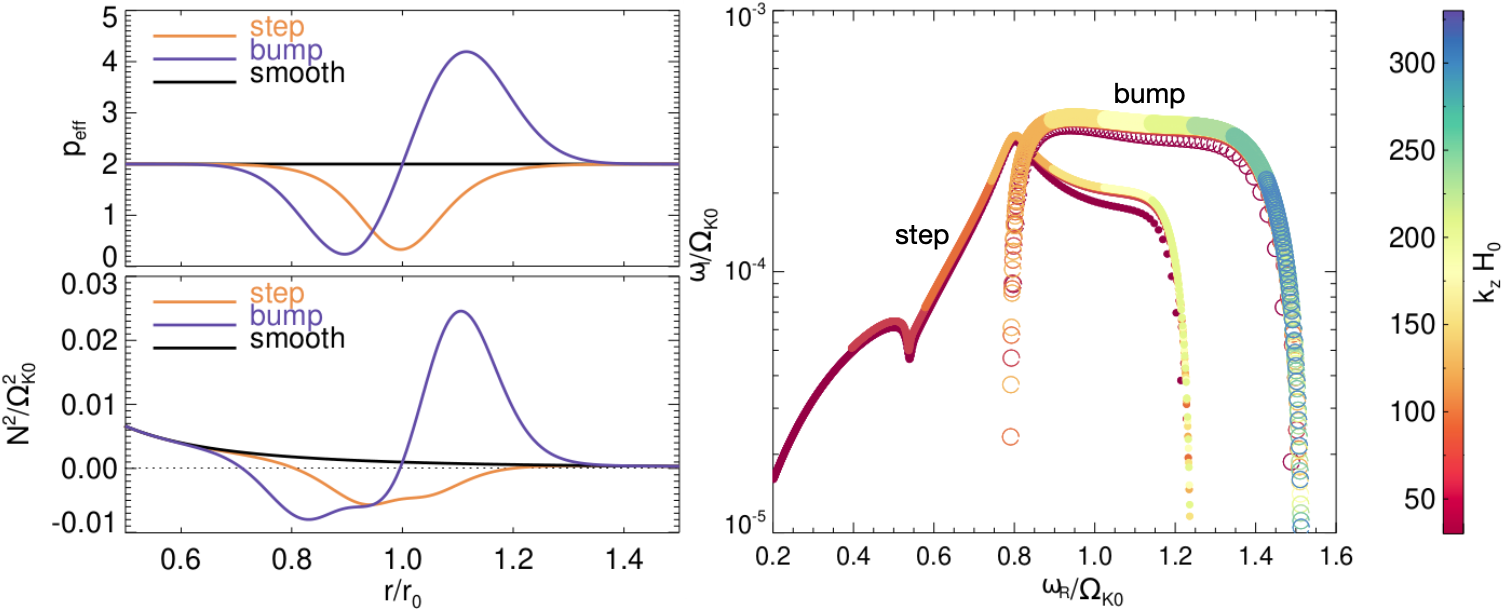}
    \caption{Illustration of the linear COS induced by a density structure in an otherwise linearly stable disk. The Left panel shows radial profiles of density $\rho$ (and the effective density slope $p_{\text{eff}}$) and squared buoyancy frequency $N^2$ of the smooth disk, and disks containing either a density bump or a density step with amplitude $A=0.4$. The right panel shows linear growth rates of COS modes for the latter two cases for different vertical wavenumbers $k_{z}$. }
    \label{fig:cos_bump_eq}
\end{figure*}

For completeness, Figure \ref{fig:eigen_var}
shows growth rates of the fastest-growing COS modes as a function of various disk parameters.
The left panel compares different values of an included $\alpha$-viscous stress\footnote{The indicated values of $\alpha$ correspond to a disk radius $r=r_0$.} (see Appendix \ref{app:visc}). Here, we also indicate the region that is expected to be resolved in our hydrodynamic simulations below (\S \ref{sec:hydrosim}). The lower boundary is set by the vertical size of the simulation domain ($\Delta z=0.5 H_{0}$), whereas the upper boundary is set by the grid resolution, assuming that 10 grid points per wavelength are required to resolve a mode. Thus, based on the linear results, we expect the COS in our simulations to be extinguished for $\alpha\gtrsim 10^{-4}$, which we will find in reasonable agreement with the simulation results presented below. The behavior of the curves in the remaining panels largely agrees with the fully local calculations presented in previous works. That is, the growth rates in the three rightmost panels are well explained by Eq. (\ref{eq:gr_loc}) using Eq. (\ref{eq:nr2}) for the buoyancy frequency (its radial average). For reference, the open circles in the second panel represent fully local growth rates computed using the method explained in \S \ref{sec:loc_glob} for the fiducial case.
It should be noted again that modes $k_{z}H_{0}\lesssim 1$ are likely to be affected by the neglect of vertical disk stratification.

\subsubsection{Disk region containing a radial density structure}\label{sec:lin_res_pbump}

\begin{figure}
\centering 
	\includegraphics[width=0.43\textwidth]{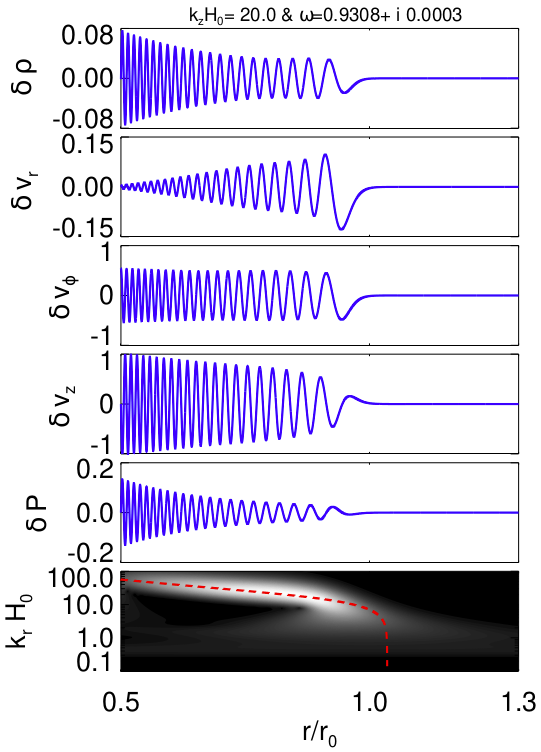}
    \caption{Hydrodynamic quantities (their real part) associated with the fastest growing COS mode with $k_{z}H_{0}=20$ in a disk containing a pressure bump (\ref{eq:bump}) with $A=0.4$, as obtained with the shooting method. The mode is excited in the region $0.7\lesssim r/r_0\lesssim 1$ where $N^2<0$ (see Figure \ref{fig:cos_bump_eq}) but penetrates into the stable region at smaller radii. The bottom panel shows the wavelet power spectrum of $\delta v_{z}$ as a function of radial wavenumber $k_{r}$.  The red dashed curve is the radial wavenumber resulting from (\ref{eq:inertial_wave}). The Lindblad resonance location where $\omega_R=\Omega$ is the location where the modes' radial wavenumber drops to zero.}
    \label{fig:cos_bump}
\end{figure}

 It is interesting to verify if the COS can operate in a more ``realistic'' disk model. By this we mean a model of a disk region with temperature and density slopes compatible with observational constraints, but containing a "special" structure that locally gives rise to the necessary temperature and density gradients to yield a negative value of $N^2$ across a certain radial domain. To this end we consider an initially convectively stable disk with $p=2$ and $q=0.75$ (cf. Figure \ref{fig:cos_crit}) on which we superimpose either a density bump 
 \begin{equation}\label{eq:bump}
\rho(r)  \to \rho(r) \times  \left[1+ A \exp\left(-\frac{\left(r-r_{0}\right)^2}{2 H_{0}^2} \right) \right],
\end{equation}
or a density step
 \begin{equation}\label{eq:step}
\rho(r)  \to \rho(r) \times  \left[1+ \frac{A}{2}\tanh{\left(\frac{r-r_{0}}{H_{0}}\right)^2} \right],
\end{equation}
to render the region unstable to the 
 COS. This is illustrated in Figure \ref{fig:cos_bump_eq} (left panel) for an amplitude $A=0.4$. The upper panel displays the effective density slope $p_{\text{eff}} =- \partial \log \rho /\partial \log r$, whereas the lower panel shows $N^2$ for the three cases of a smooth disk, and a disk containing a density bump and a density step, respectively.
 A linear analysis adopting the latter two background disk structures yields growing modes for $k_{z} H_{0}\gtrsim 5$. For increasing $k_{z}$ bands of growing modes are found which become increasingly narrow and concentrated around the largest Keplerian frequencies within the unstable region. The fastest growing modes have their LR close to the outer boundary of the unstable region. However, note that even for large $k_{z} H_{0} \gg 1$ the fastest growing modes possess growth rates which are substantially smaller than in the smooth disk case discussed above. 
 
It is interesting to note that in the present case most\footnote{We find some modes with very small growth rates $\omega_{I}\lesssim 5\cdot 10^{-5}$ that traverse the entire domain. } of the growing linear modes in this problem resemble to some extent, resonantly forced (spiral) density waves (e.g., \citealt{shu1984}) since they similarly appear to be excited within a narrow, unstable region and propagate away from it with radially decreasing wavelength. As in the case of the smooth disk, here we also find only modes radially inward of their Lindblad resonances. In Figure \ref{fig:cos_bump} we show the fastest growing mode with $k_{z} H_{0}=20$.

\section{Hydrodynamic Simulations}\label{sec:hydrosim}

\subsection{Numerical setup}\label{sec:numerics}

We use FARGO3D\footnote{\url{http://fargo.in2p3.fr}} \citep{fargo3d,llambay2019}  to perform hydrodynamic simulations in which we integrate Eqs. (\ref{eq:contrhog})---(\ref{eq:contp}) forward in time.  
As our disk is vertically unstratified, we apply periodic boundary conditions to all quantities in the azimuthal and vertical directions.
Radial
boundary conditions are such that equilibrium values of gas density and azimuthal velocity are extrapolated into the ghost zones, while gas radial and vertical velocities are set to zero at the boundaries. Simulations are carried out in cylindrical coordinates as defined in \S \ref{sec:model}. Computational units are such that $r_{0}=M_{*}=G=1$.
We adopt $h_{0}=H_{0}/r_{0} =0.1$ in all runs. For our fiducial setup
 the numerical grid covers $0.5\leq r/r_{0}\leq 1.5$ and $-0.25H_0\leq z \leq +0.25 H_0$, and in 3D simulations $0\leq \varphi \leq 2 \pi$.
The grid resolution is $n_{r}\times n_{z} \times n_{\phi} = 2000 \times 100 \times 628$. Thus, the radial and vertical resolutions are $200/H_{0}$, whereas the azimuthal resolution $\sim 10/H_0$ is significantly lower for reasons of computational costs\footnote{The typical size of a single 3D simulation is $\approx 600 $Gb.}, and since we expect structures to form in our simulations to be predominantly axisymmetric or elongated in azimuthal direction. 
3D simulations are carried out on a GPU cluster, which is necessary due to the high spatial resolution. All 3D simulations cover 1,000 reference orbits.

\subsection{Diagnostics}\label{sec:diagnostics}

Following previous studies \citep[e.g.][]{lehmann2022}, we describe the radial and vertical turbulent angular momentum transport in 3D simulations via the dimensionless quantities
\begin{equation}\label{eq:alpha}
 \alpha_{r}(t) = \frac{\langle \rho v_{r} \delta v_{\varphi} \rangle_{r\varphi z}}{\langle P \rangle_{r \varphi z}}
\end{equation}
and
\begin{equation}\label{eq:alpha_z}
 \alpha_z(t) = \frac{\langle  \rho v_{z}  \delta v_{\varphi} \rangle_{r\varphi z}}{\langle P \rangle_{r \varphi z}},
\end{equation}
respectively, where the brackets denote averaging over spatial dimensions as indicated in the subscript, and where $\delta v_{\varphi}$ is the azimuthal velocity deviation from its ground state value. 

Furthermore, we define the deviation $\Delta X$ of a quantity $X(r,\varphi,z)$ from its vertical-azimuthal average across a cylindrical annulus at radius $r$ via
\begin{equation}\label{eq:pert}
    \Delta X \equiv  X - \frac{ \int \limits_{0}^{2 \pi}   \int \limits_{z_{\text{min}}}^{z_{\text{max}}} X  r \mathrm{d}\varphi \mathrm{d}z   }{\int \limits_{0}^{2 \pi}   \int \limits_{z_{\text{min}}}^{z_{\text{max}}}  r \mathrm{d}\varphi \mathrm{d}z }.
\end{equation}
If not stated otherwise, our radial diagnostic domain is defined by $r_{\text{min}}/r_0=0.8$ and $r_{\text{max}}/r_0=1.2$.

\subsection{Results of 2D axisymmetric  simulations}\label{sec:sim_2d}

 In this section we present the results of 2D axisymmetric  (radial-vertical) simulations of the nonlinear saturation of the COS. In \S \ref{sec:smooth_2d} we consider simulations of a smooth power-law disk with $N^2<0$ for all radii. Next, in \S \ref{sec:struct_2d} we consider a disk containing a pressure bump which locally gives rise to the COS in an otherwise stable disk. The different values of $p$ and $q$ adopted throughout our simulations are shown in Figure \ref{fig:cos_crit} for illustration.

\subsubsection{Nonlinear saturation of the COS in a smooth disk}\label{sec:smooth_2d}

\begin{figure*}
\centering 
	\includegraphics[width= 0.9\textwidth]{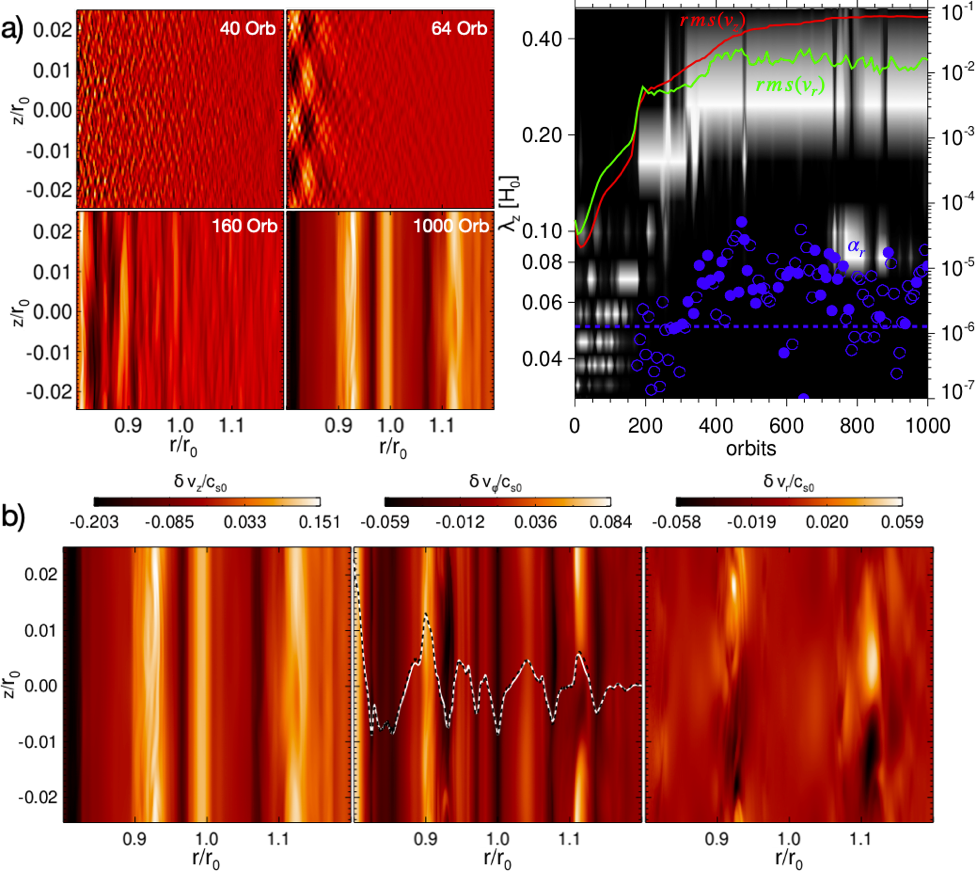}
    \caption{\textbf{a)}: The left panels show snapshots of the vertical velocity in the course of the axisymmetric nonlinear saturation of the COS, where brighter (darker) colors correspond to larger (smaller) values of $\delta v_z$. Note that the color scaling of the bottom right frame at $1000$ orbits is the same as in the left panel in the bottom frame \textbf{b)}. The right frame shows the power spectral density of the radial velocity in vertical wavelength space, and illustrates the increased vertical elongation of structures, eventually saturating into elevator flows and zonal flows with almost vanishing vertical structure. The curves represent rms radial and vertical velocities as indicated. The circles and the dashed line represent $\alpha_{r}(t)$ and its time-average, respectively. See the text for details \textbf{b)}: Saturated state of the COS after 1,000 orbital periods, exhibiting persistent  elevator flows (left panel) and zonal flows (middle panel). The dashed curves in the middle panel illustrate the geostrophic balance of the displayed zonal flows as explained in the text.}
    \label{fig:cos_satur}
\end{figure*}

\begin{figure*}
\centering 
	\includegraphics[width=\textwidth]{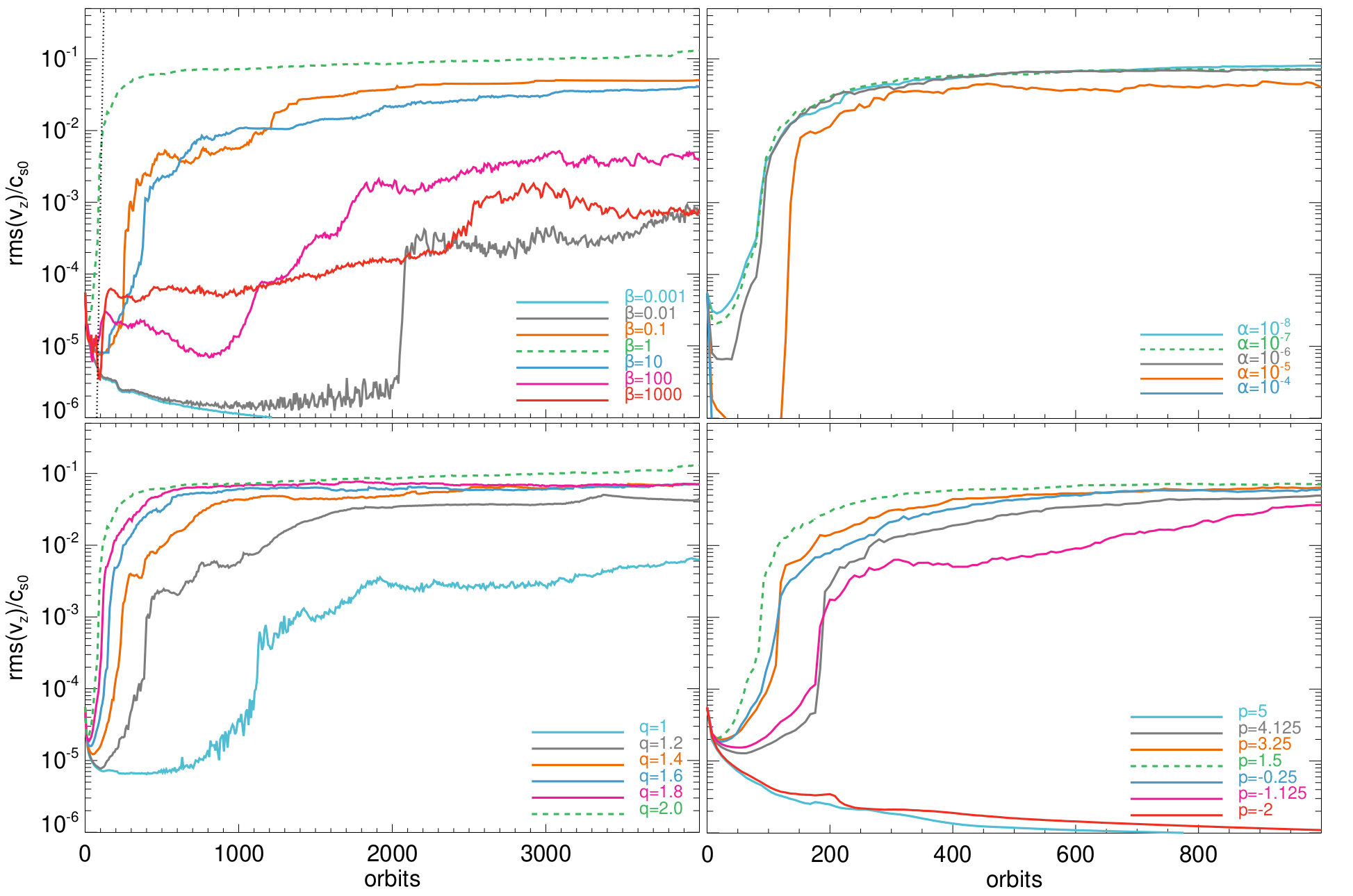}
    \caption{rms vertical velocities in 2D axisymmetric hydrodynamic simulations. Compared are different cooling times $\beta$ (upper left), initial temperature slopes $q$ (lower left), initial density slopes $p$ (lower right) and values of the dimensionless viscosity $\alpha$ defined via Eqs. (\ref{eq:reynolds}) and (\ref{eq:stresstensor}) at $r=r_0$. The dashed line near $100$ orbits in the upper left panel represents the maximum growth rate predicted by the linear radially global analysis above. It matches well with the largest linear growth rate seen in the simulation for the same disk parameters. Note that the COS in simulations with $\alpha \gtrsim 10^{-4}$ is fully damped for at least 10,000 orbits (not shown).}
    \label{fig:cos_var}
\end{figure*}

Figure \ref{fig:cos_satur} illustrates the nonlinear saturation of the COS in our fiducial axisymmetric simulation. The upper left frame shows the evolution of initially small scale, quasi-local modes into radially global modes. The latter eventually develop persistent flow structures. The upper right frame shows the evolution of the power spectral density of the radial velocity field (contours), averaged over the radial diagnostic domain $0.8\leq r/r_{0} \leq 1.2$, as a function of vertical wavelength $\lambda_z$. Generally, we find that the prevalent vertical length scale of flow structures increases with time. Note that in the saturated state the prevalent vertical wavelength of the radial velocity, which should represent secondary parasitic modes (\citealt{latter2016};~\hyperlinkcite{teed2021}{TL21}) is $L_{z}/2$. This is consistent with our numerical convergence analysis, presented in Appendix \ref{app:convergence}, which shows that smaller vertical domain sizes than our fiducial one result in a weakened nonlinear saturation. 
The over-plotted solid curves are the rms vertical (green) and radial (cyan) velocities. As explained in \hyperlinkcite{teed2021}{TL21}, the radial (vertical) rms velocities represent the power in the primary COS (secondary parasitic) modes. The onset of parasitic modes eventually controls the nonlinear saturation amplitude of the COS. 

Furthermore, the circles represent the turbulent radial angular momentum flux $\alpha_{r}$ computed via (\ref{eq:alpha}), such that open (solid) circles denote negative (positive) values. The dashed line denotes its \emph{negative} time-averaged (over the last 400 orbits) value $\langle \alpha_{r} \rangle_{t}\approx -1.3 \cdot 10^{-6}$, and is roughly in agreement with corresponding values reported by \hyperlinkcite{teed2021}{TL21} (their Figure 17) for $R\approx 0.05$. Note that since our scaling length is $H$, rather than $L_{z}$ (as in \hyperlinkcite{teed2021}{TL21}), our value is to be multiplied by a factor of 4 for comparison.

The bottom frames in Figure \ref{fig:cos_satur} show snapshots of the vertical, azimuthal and radial velocity field components, respectively, in the saturated state of the simulation. The left and middle panels show the presence of persistent elevator flows and zonal flows, respectively, which constitute radial patches of alternating deviations of the corresponding velocity components (see \hyperlinkcite{teed2021}{TL21} for details on these structures). Zonal flows are accompanied by significant undulations in the radial pressure gradient, the latter being required to establish geostrophic balance in these flow structures. This is verified by the over-plotted curves of $2 \Omega \delta v_{\varphi} $ and $1/\rho \,\partial_{r} \delta P$, which are practically identical, thereby demonstrating geostrophic balance between the Coriolis force and the radial pressure gradient. 
Note that here, pressure variations $\delta P$ are caused mainly due to variations in the density, as temperature variations are found to be subdominant.

In Figure \ref{fig:cos_var} we plot rms vertical velocities illustrating the dependence of the nonlinear saturation of the COS on various physical simulation parameters. These are the cooling time $\beta$ (upper left panel), the input viscosity $\alpha$ defined via Eq. (\ref{eq:reynolds}) (upper right panel), the initial radial density slope $p$ (lower left panel) and the initial radial temperature slope $q$ (lower right panel). In all panels of Figure \ref{fig:cos_var} the dashed curve represents our fiducial simulation. Note the different simulation times in the left and right panels . The fiducial values of $\beta$ and $p$ are chosen so as to maximize linear growth rates and the saturation level of the COS reached within $1,000$ orbits.
The adopted viscosity $\alpha = 10^{-7}$ is roughly the largest possible value which does not notably affect the evolution of the instability. All simulations are initialized with low amplitude white noise $\sim 10^{-3}c_{s0}$ in all velocity components.

For $\alpha=10^{-7}$ we have $R_{\text{crit}}=0.0032$ by Eq. (\ref{eq:ncrit}).
The simulations with varying density slope and temperate slope which develop COS\footnote{For the simulations with $p=-2$ and $p=5$ the buoyancy frequency $N^2$ vanishes identically. Therefore,  these simulations merely serve as a consistency check.} yield values in the range $0.037\lesssim R\lesssim 0.085$ and $0.008\lesssim R \lesssim 0.085$ within the diagnostic domain, respectively, such that in all of these simulations persistent zonal flows are expected to develop, at least after sufficient simulation time (see \S \ref{sec:prelim}). This is found to be largely the case, except for simulations with large cooling times $\beta\gtrsim 100$. We suspect that in the latter simulations initially another double diffusive instability develops, making use of viscous diffusion (see Appendix \ref{sec:double_diffusive} for details). However, after roughly 2,000 orbits we find the emergence of elevator flows, indicating that the double diffusive instability eventually gets supplanted by the COS. In the simulations presented here the former instability is weak due to the small values of $\alpha$ adopted. 
We hypothesize that after sufficient integration time the COS eventually saturates to similar levels in all simulations shown in the upper left panel. 

 We also performed additional simulations using wave damping boundary conditions (not shown). In these simulations we find a very similar nonlinear saturation, albeit slightly delayed (by 100-200 orbits) compared to the simulations shown in Figure \ref{fig:cos_var}. However, we find that the COS is almost entirely suppressed (for at least 10,000 orbits) in simulations using damping boundary conditions with $p=4.125$ and $p=-1.125$, and also in simulations with cooling times $\beta\gtrsim 10$ (with final rms vertical velocity $\sim 10^{-6}$ which keeps decreasing). This is an interesting result, as it suggests that the COS developing in our simulations is indeed a radially global, rather than a local instability.

\begin{figure*}
\centering 
	\includegraphics[width=\textwidth]{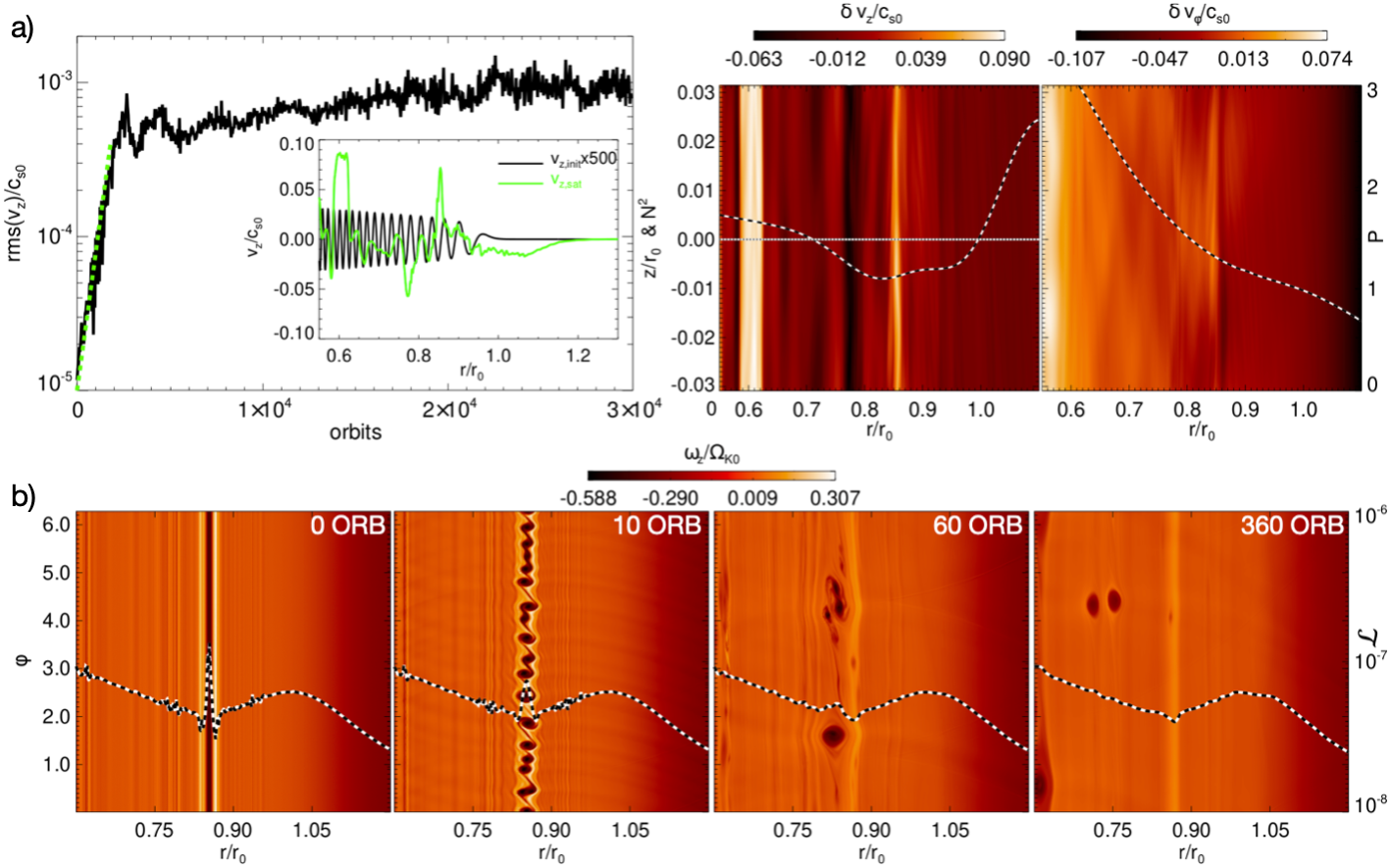}
    \caption{\textbf{a):} Nonlinear saturation of the COS in the vicinity of a pressure bump. The upper left frame shows the evolution of the rms vertical velocity. The dashed line indicates the linear growth predicted by our linear analysis. The insert figure displays initial and final  vertical velocities (at $z=0$). 
 The upper right frame shows contours of the vertical and azimuthal velocity perturbations. The over-plotted curves are the initial radial buoyancy and pressure profiles. \textbf{b)}: Contours of the vorticity (\ref{eq:vort}) defined below of a 3D simulation for which we used the axisymmeteric saturation depicted in the top panels as initial state. The dashed curves represent the vortencity $\mathcal{L}$, defined in Eq. (\ref{eq:vorten}).}
    \label{fig:cos_bump_seed}
\end{figure*}

\subsubsection{Nonlinear saturation of the COS in the vicinity of a pressure bump}\label{sec:struct_2d}

In \S \ref{sec:lin_res_pbump} we conducted a linear analysis of a disk region containing density structures which give rise to the COS via a local modification of $N^2(r)$ in an otherwise stable disk, corresponding to the left most circle in Figure \ref{fig:cos_crit}.
Figure \ref{fig:cos_bump_seed} illustrates the nonlinear evolution of the COS in the vicinity of a pressure bump (the same as in Figure \ref{fig:cos_bump_eq}). 
In this simulation, we seeded the linear eigenfunction with $k_{z} H_{0}=20$ (Figure \ref{fig:cos_bump}) with a small initial amplitude.

A similar outcome of the nonlinear saturation is obtained by seeding low amplitude white noise as done before. 
However, seeding the fastest growing mode significantly speeds up the evolution. 
Moreover, we can test its growth rate against the linear analysis presented above.
The vertical domain for this simulation is set to be $\Delta z =\pi/5 H_{0}$, such that the seeded mode fits exactly two times in the vertical domain. The black curve in the upper left frame displays the rms vertical velocity, with a linear growth phase of about 1,000 orbits, and the subsequent nonlinear saturation. The dashed green line represents the linear growth rate as predicted by our linear analysis, and which is indicated also in Figure \ref{fig:cos_bump}. 
The insert figure shows the initial and final radial profile of the vertical velocity (at $z=0$). 
Thus, even though only the region $0.7\lesssim r/r_0 \lesssim 1$ is unstable to the COS, excited waves penetrate significantly into the stable region at smaller radii. This is further illustrated in the upper right frames, which show contour plots of the vertical and azimuthal velocity perturbations after 30,000 orbits. The over-plotted curves are the buoyancy and pressure profiles $N^2$ and $P$, respectively.

\subsection{Results of 3D simulations}\label{sec:3dsims}

We now shift our focus to our 3D simulations. Given the significant computational resources needed, we present a concise qualitative summary instead of a comprehensive parameter exploration. 
\begin{figure*}
\centering 
	\includegraphics[width=\textwidth]{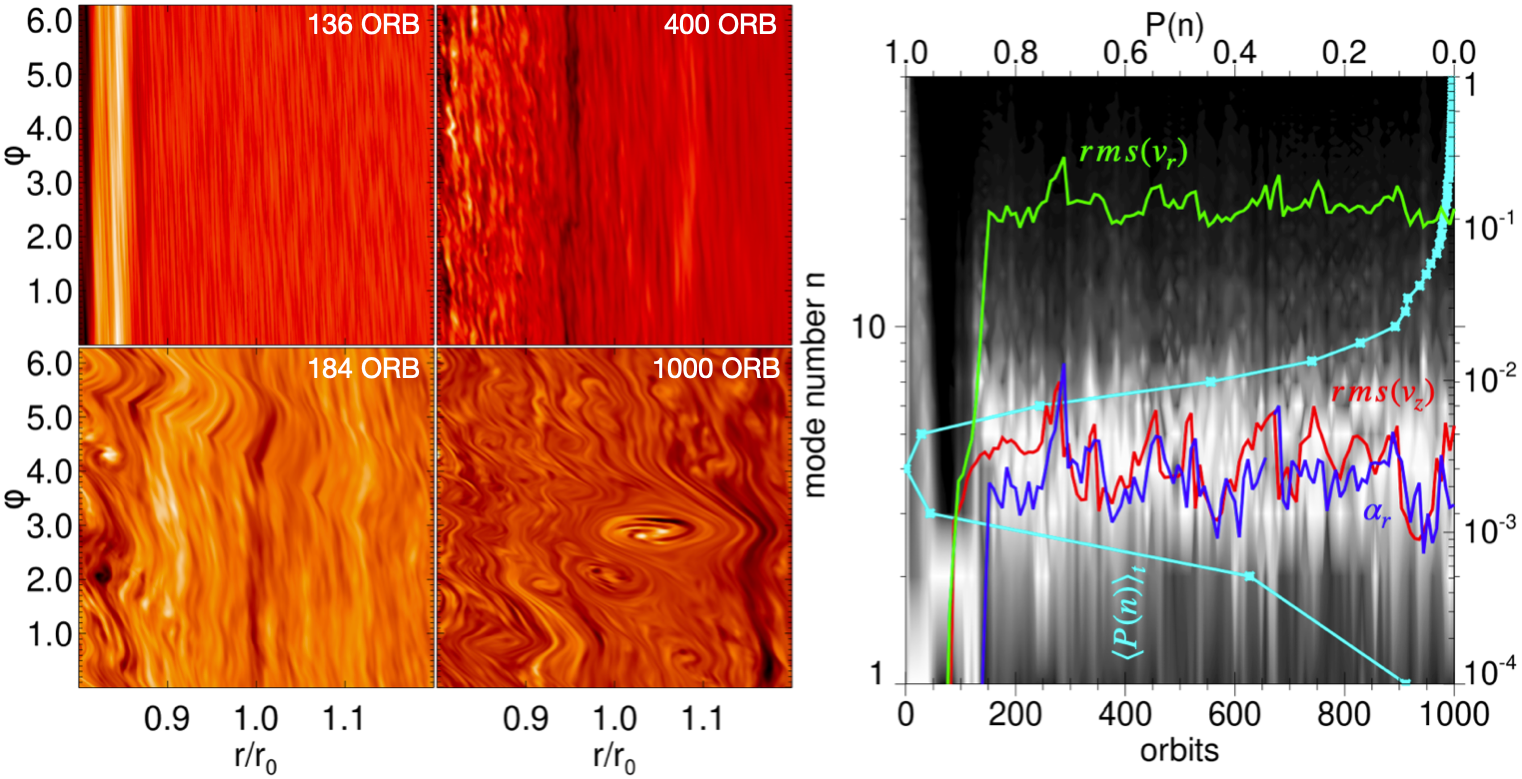}
    \caption{Nonlinear evolution of the COS in the fiducial 3D simulation. The left panel shows snapshots of the vertically averaged vertical velocity. In the right panel the contour plot represents the time evolution of the azimuthal power spectral density of the radial velocity (planar cut at $z=0$). The curve labeled "$\langle P(n) \rangle_{t}$" is the time-averaged normalized function (\ref{eq:powerspec}), corresponding to the left plot-axis. The remaining curves are the time evolution of the rms vertical velocity, rms radial velocity  and $\alpha_{r}$, corresponding to the right plot-axis.}
    \label{fig:cos_evol_3d}
\end{figure*}
In particular, we perform 3D simulations of smooth disks (\S \ref{sec:smooth_3D}) where we consider three different values of the cooling time $\beta=0.1$, $\beta=1$ and $\beta=10$, as well as four different values of the initial density slope $p=-0.25$, $p=0.5$, $p=1.5$ and $p=2.5$. The latter values are also shown in Figure \ref{fig:cos_crit}. In the simulations adopting different $p$ we slightly adjust the temperature slope $q$, such that the radial profile of $N^2$ is the same in all simulations. Thus, essentially $q\approx 2$ in these simulations. In addition, we consider one semi-3D simulation in the vicinity of a pressure bump, which we discuss first in \S \ref{sec:semi3d}

\subsubsection{Semi-3D simulation of the COS in the vicinity of a pressure bump}\label{sec:semi3d}

Before we describe our main results below, we briefly discuss one particular 3D simulation illustrated in the bottom panels of Figure \ref{fig:cos_bump_seed}. The latter shows the contours of the vorticity of a 3D simulation that adopted the axisymmetric nonlinear saturation shown in the top panels of the exact figure as the initial state. In particular, the zonal flow formed around $r\approx 0.9$ becomes unstable to the Rossby wave instability \citep{lovelace1999} and spawns several small vortices. 
This is consistent with the occurrence of a maximum in the vortencity $\mathcal{L}$ (over-plotted dashed curve in all panels) corresponding to the zonal flow, where
\begin{equation}\label{eq:vorten}
\mathcal{L} = \frac{\Sigma}{2 \omega_{z}}  \left(\frac{\Pi}{\Sigma^{\gamma}}\right)^{2/\gamma} 
\end{equation}
with the vertically integrate pressure $\Pi$.
The vortices eventually merge and strengthen, and the vortensity maximum disappears. The vortices then excite (weak) spiral density waves (cf \S \ref{sec:spirals}) and migrate radially inward. The corresponding angular momentum flux is minimal, though, taking values $\alpha_{r}\sim 10^{-5}$ (not shown). 
Note that this simulation's non-axisymmetric outcome is likely somewhat flawed due to the artificial 'switching on' of the azimuthal dimension only after 30,000 orbits. It is unclear if the zonal flow would have attained the same strength if the simulation had been 3D from the outset.


\subsubsection{Nonlinear saturation of the COS in a smooth disk}\label{sec:smooth_3D}

Figure \ref{fig:cos_evol_3d} illustrates the nonlinear saturation of the COS in our fiducial 3D simulation ($p=1.5$, $q=2$). Compared to Figure \ref{fig:cos_satur} we show here the evolution of the vertically averaged vertical velocity (left panels) in the $r\varphi$-plane. The vertical structure of the flow is similar to our 2D simulations (\ref{sec:sim_2d}), but no notable \emph{persistent} zonal flows and elevator flows form, as these are rapidly turned into vortices. 
\begin{figure*}
\centering 
	\includegraphics[width=\textwidth]{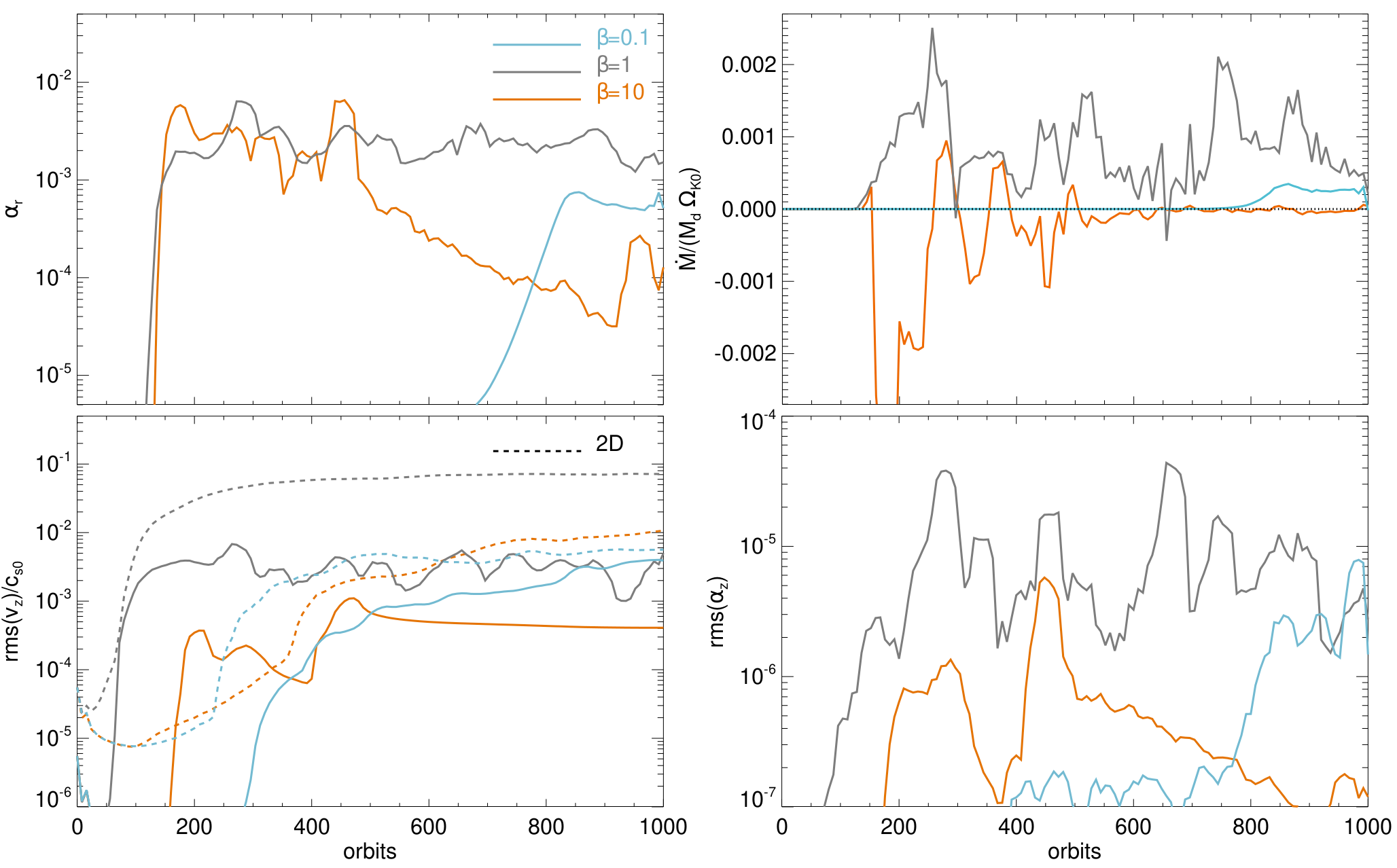}
    \caption{ Radial transport of angular momentum (top left), rms vertical velocities (bottom left), radial mass accretion rate (top right), and rms vertical transport of angular momentum (bottom right) in 3D simulations with varying cooling time $\beta$. The quantity $M_{d}$ denotes the initial mass contained within the diagnostic domain in each simulation.}
    \label{fig:3D_compare_beta}
\end{figure*}
The right panel shows the normalized azimuthal power spectrum of the mid-plane radial velocity field
\begin{equation}\label{eq:powerspec}
    P(n) \sim \left| \, \, \int \limits_{0}^{2 \pi} \int \limits_{r_{\text{min}}}^{r_{\text{max}}}  \mathrm{d}\varphi \mathrm{d}r  e^{i n \varphi} \, \delta v_r(z=0)\right|^2, 
\end{equation}
where $n$ is a positive integer denoting the azimuthal mode number.
The curves describe the rms vertical (red) and radial (green) velocities, as well as $\alpha_r(t)$ (blue). Most power is located in modes with azimuthal mode numbers $n=3-5$, owing to the excitation of spiral waves by vortices, which are responsible for the radial angular momentum transport, measured through $\alpha_r$ (see \S \ref{sec:spirals}).
It is noteworthy that the magnitude of the rms radial and vertical velocities is reversed as compared with our 2D simulations. This holds true for all simulations we conducted in this study.  We speculate that this difference is related to the presence of large-scale vortices in 3D simulations, in contrast to persistent elevator flows in 2D simulations.

\begin{figure*}
\centering 
	\includegraphics[width=\textwidth]{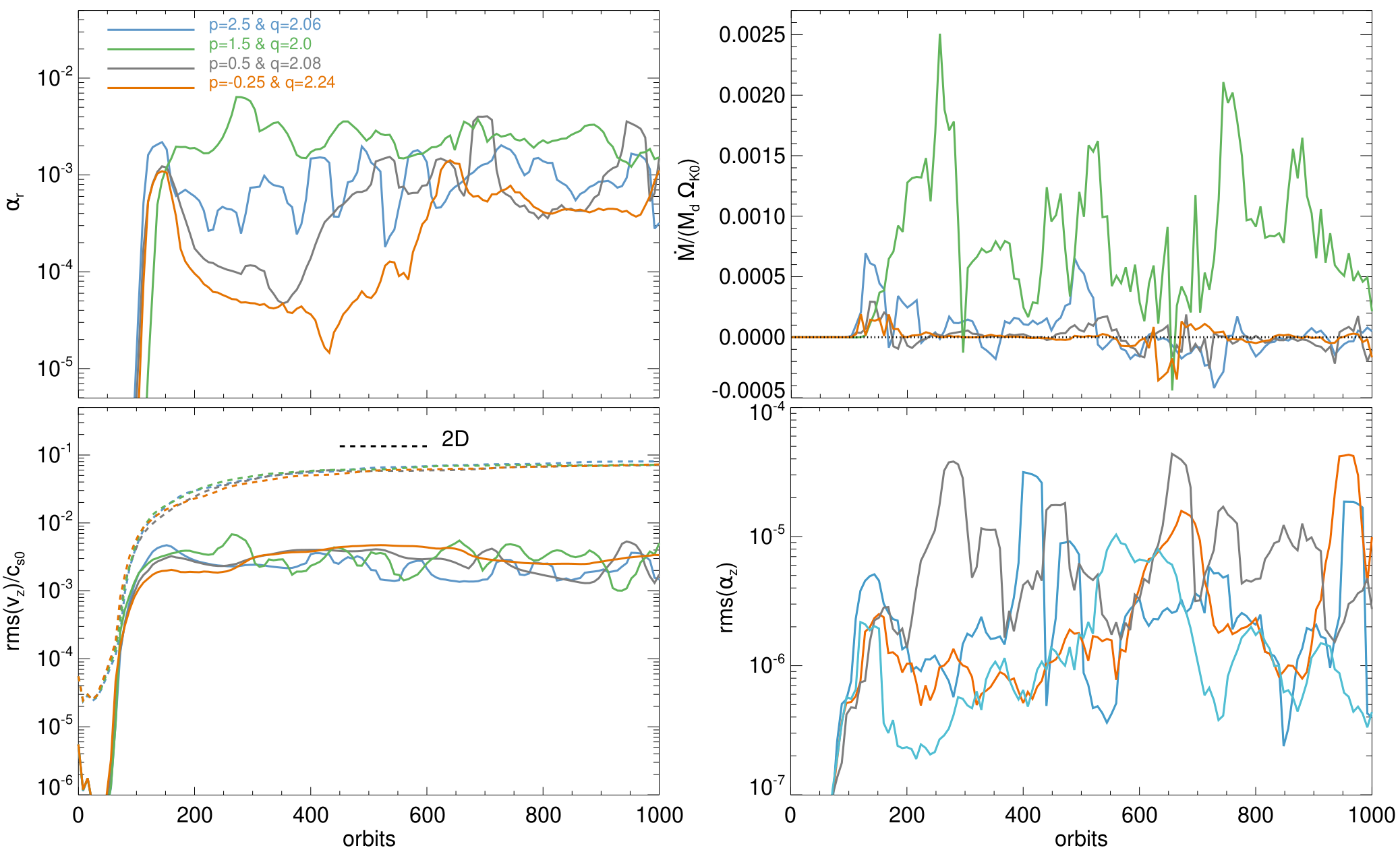}
    \caption{Same as Figure \ref{fig:3D_compare_beta} but now for simulations with varying initial density slope $p$.}
    \label{fig:3D_compare_pslop}
\end{figure*}

In Figures \ref{fig:3D_compare_beta} and \ref{fig:3D_compare_pslop} we show the time evolution of  $\alpha_r$ (upper left panels), the rms vertical velocities (lower left panels), the radial mass accretion rate 
\begin{equation}\label{eq:mdot}
    \dot{M} \equiv \int\limits_{0}^{2\pi}\int\limits_{z_{\text{min}}}^{z_{\text{max}}} r \delta v_{r} \rho \mathrm{d }\varphi \mathrm{d}z
\end{equation}
(upper right panels), and $\text{rms}(\alpha_{z})$ (lower left panels), for different cooling times $\beta$, and density slopes $p$, respectively. 
Our measured $\alpha_{r}$-values are in the range $10^{-4}- 10^{-3}$, roughly consistent with values obtained from local compressible shearing box simulations \citep{lyra2014}.

\subsubsection{Mass and angular momentum transport}\label{sec:transport}

In this section, we study the radial transport of mass and angular momentum resulting from the COS in 3D simulations in more detail. Note that the negative (time-averaged) angular momentum flux induced by the nonlinear COS in our axisymmetric simulations is minimal (cf. \S \ref{sec:smooth_2d}), such that the disk's radial density structure is not altered to any significant extent (not shown). This is consistent with the results of \hyperlinkcite{teed2021}{TL21}.
Despite its small magnitude though, the axisymmetric radial angular momentum flux is the driving element for the formation of zonal flows, and hence vortices (see \S 6 of \hyperlinkcite{teed2021}{TL21} for details). 
As mentioned above, in our 3D simulations we generally find a significant outward radial transport of angular momentum. 
Interestingly, in most cases this goes along with an \emph{outward} radial transport of mass. For the case $p=1.5$ this is true at least for cooling times $\beta=0.1$ and $\beta=1$. 
Indeed, using Eq. (\ref{eq:mdot}) we find time-averaged accretion rates $\langle \dot{M}\rangle_{t}=-8.3\cdot 10^{-10}$, $1.6\cdot 10^{-9}$, $2.3\cdot 10^{-7}$, $1.5\cdot 10^{-8}$, for simulations adopting the initial slopes $p=-0.25,0.5,1.5,2.5$, in units $M_{*} \, \Omega_{K0}$. We also find $\langle \dot{M}\rangle_{t}=1.6\cdot 10^{-8}$, $-8.9\cdot 10^{-8}$, for simulations adopting the cooling times $\beta=0.1,10$ (and $p=1.5$).
The finding of a positive accretion rate (\ref{eq:mdot}) is contrary to the case of a laminar viscous accretion disk with a \emph{constant} kinematic viscosity.
Nevertheless, our results are consistent with axisymmetric viscous accretion disk theory \citep{lynden1974} as we show in the following.

Using (\ref{eq:mdot}) we can define the effective accretion velocity, as obtained from our simulations, through
\begin{equation}\label{eq:vacc_1}
 v_{r,\text{sim}} \equiv \frac{\dot{M}}{2 \pi r \Sigma}   ,
\end{equation}
with the surface mass density
\begin{equation}
    \Sigma = \frac{1}{2\pi}\int\limits_{0}^{2\pi}\int\limits_{z_{\text{min}}}^{z_{\text{max}}}  \rho \mathrm{d }\varphi \mathrm{d}z.
\end{equation}
For an axisymmetric laminar, viscous Keplerian accretion disk we have \citep{lynden1974}
\begin{equation}\label{eq:vacc_2}
    v_{r,\text{axi}} \equiv -\frac{3}{\sqrt{r}\Sigma}\partial_{r}\left(\overline{\nu}\Sigma \sqrt{r} \right),
\end{equation}
where in the current situation (i.e. for an $\alpha$-viscous prescription of a turbulent, non-axisymmetric accretion disk) we define the axisymmetric kinematic shear viscosity via
\begin{equation}\label{eq:viscosity}
\overline{\nu}(r,t)\equiv \alpha_{r}(r,t) \, \frac{c_{s0}^2}{\OmK}
\end{equation}
with (cf \S \ref{sec:diagnostics})
\begin{equation}\label{eq:alpha_rt}
\alpha_{r}(r,t) = \frac{\langle \rho v_{r} \,\delta v_{\varphi} \rangle_{\varphi z}}{\langle P \rangle_{ \varphi z}}.
\end{equation}

In particular, $\alpha_r(r,t)$ and thus $\overline{\nu}$ may vary with radius (and time), which explains our results, as shown below. Note that the angular frequency $\Omega\approx \OmK$ and the sound speed $c_{s}^2$ given through (\ref{eq:soundspeed}) and (\ref{eq:T_equ}) do not change significantly during our simulations.
Using (\ref{eq:vacc_1}) and (\ref{eq:vacc_2}) we find (for an axisymmetric laminar, viscous Keplerian accretion disk)
\begin{equation}\label{eq:mdot2}
    \dot{M} \sim \left[m(t)-\frac{1}{2}\right] \left(\frac{r}{r_0}\right)^{-m(t)}
\end{equation}
with
\begin{equation}
m(t)\equiv a(t)+p(t)+3 q/2 -3,
\end{equation}
 where we assumed 
\begin{equation}
    \alpha_{r}(r,t) \sim \left(\frac{r}{r_0}\right)^{-a(t)}
\end{equation}
for simplicity. Note that $a(t)$ and $p(t)$, and hence also $m(t)$ vary with time in our simulations. In contrast, as mentioned above, $q$ remains essentially unchanged. 
In actuality, the density profile substantially deviates from a simple power law at later stages of our simulations (cf. Figures \ref{fig:cos_accretion_1} and \ref{fig:disc_evol} below), but for simplicity we ignore this here.
In the current approximation a steady state density profile (such that the accretion rate is radially constant) strictly requires $m(t)=0$.
Furthermore, the pre-factor $[m(t)-\frac{1}{2}]$ determines the sign and overall magnitude of the accretion rate.

Figure \ref{fig:cos_accretion_1} shows time- azimuth- and height-averaged radial profiles of $\alpha_{r}$ (top panel) and $\rho$ (bottom panel) for simulations adopting different initial  density slopes $p$. The dashed curves are power law profiles that have been fitted to the solid curves.
The corresponding averaged power law slopes are $\overline{a(t)}=0.74,1.40,3.68,0.69$ and $\overline{p(t)}=-0.68,-0.23,-0.32,0.68$ for simulations adopting the initial slopes $p=-0.25,0.5,1.5,2.5$.
Using these values we find $\overline{m(t)}=0.41,1.29,3.36,1.46$. 
 Thus, once the COS saturates, none of our simulated disks is predicted to remain in steady state, which is indeed what we find in our simulations. Note that a larger absolute value of $m$ corresponds to an increasingly radially varying accretion rate, which should result in an increasingly large modification of the radial density profile. Indeed, the above $m$-values are qualitatively compatible with the amount of radial mass-redistribution in our simulations, which is shown in Figure \ref{fig:disc_evol} (right frames). 
The left frames show simulations with different cooling times. As seen in the panels for $N^2$, large portions of the simulation region are stabilized with respect to the COS at later stages in some of the simulations. Moreover, almost all simulations form at least one pressure bump, such that $\partial_{r} \langle P \rangle_{\varphi z} =0$.

The aforementioned values of $\overline{m(t)}$ are also in qualitative agreement with the values of the time-averaged accretion rate $\langle\dot{M}\rangle_{t}$ given above. Indeed, using the values of $m(t)$ just derived, only the simulation with $p=-0.25$ is predicted to have a negative accretion rate, whereas the other simulations are predicted to yield a positive accretion rate. Moreover, the largest value is predicted for the case $p=1.5$, followed by the case $p=2.5$, etc. , which is in agreement with the values obtained from our simulations.
Similarly, for the simulation with $p=1.5$ and the longer cooling time $\beta=10$ we find $\overline{a(t)}=-2.03$, $\overline{p(t)}=0.93$ and $\overline{m(t)}=-1.10$, 
which is also in qualitative agreement with the measured accretion rate, which is negative.

Finally, Figure \ref{fig:cos_accretion_2} shows the time evolution of the radial accretion velocity, obtained from Eqs. (\ref{eq:vacc_1}) and (\ref{eq:vacc_2}). Based on the degree of agreement between the two quantities, we again conclude that the mass transport can \emph{qualitatively} be described by an axisymmetric laminar viscous accretion disk model.

\begin{figure}
\centering 
	\includegraphics[width=0.5 \textwidth]{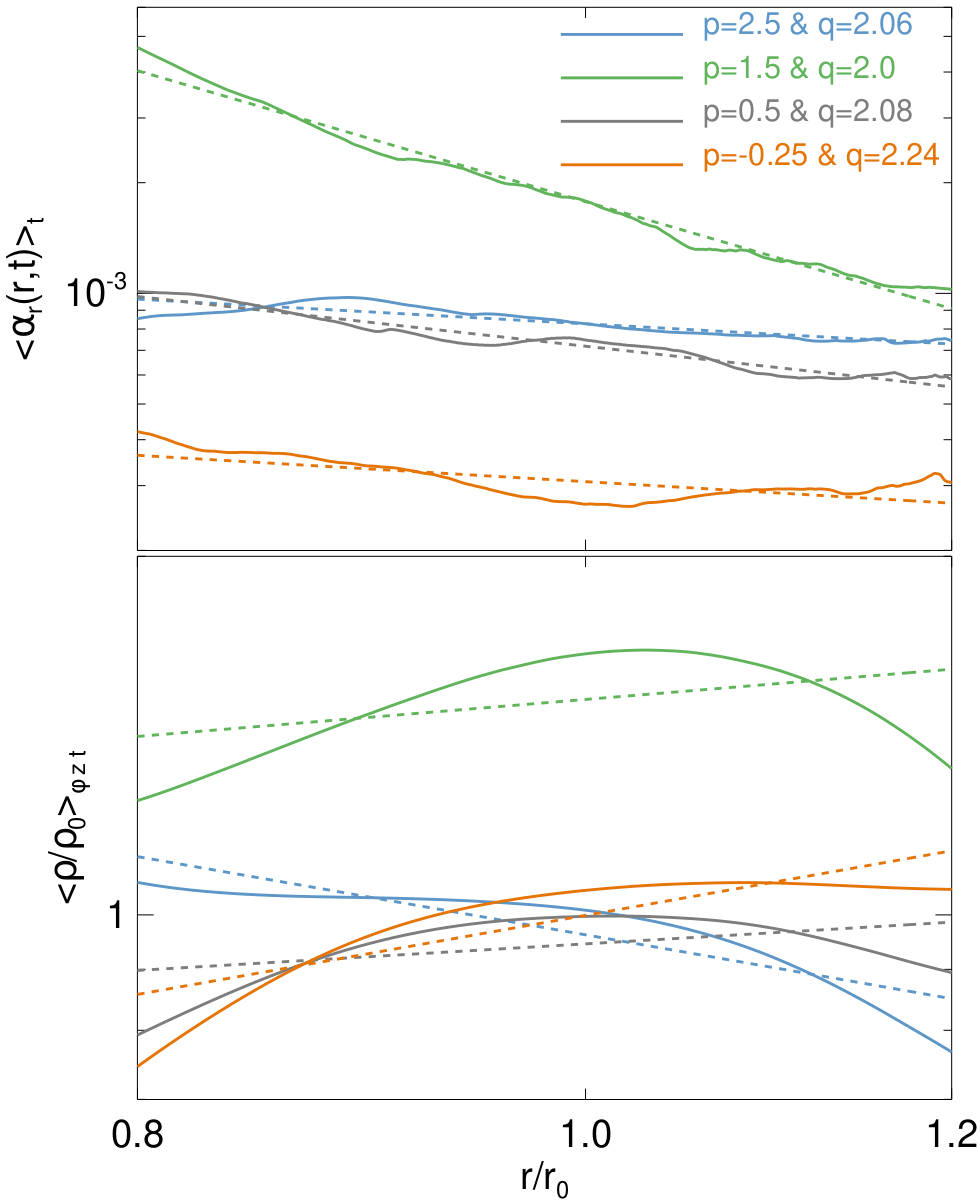}
    \caption{The top and bottom panels show radial profiles (solid curves) of the time-averaged $\alpha_r(r,t)$ as defined in (\ref{eq:alpha_rt}) and mass density $\rho$ (averaged over time, azimuth and height), respectively, for simulations adopting different initial density slopes $p$. The dashed lines are corresponding linear fits. Note that the plots are double logarithmic.}
    \label{fig:cos_accretion_1}
\end{figure}

\begin{figure}
\centering 
	\includegraphics[width=0.5 \textwidth]{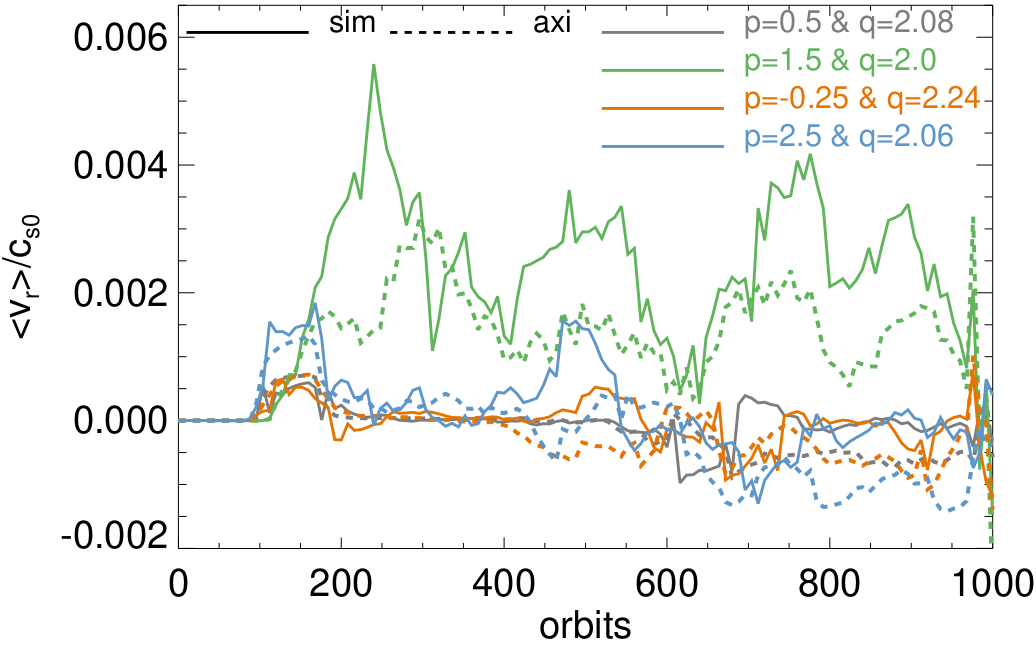}
    \caption{Testing the applicability of axisymmetric viscous accretion disk theory to our 3D simulations of the COS. We compare the two different expressions for the radial accretion velocity (\ref{eq:vacc_1}) and (\ref{eq:vacc_2}), labeled "sim" and "axi", respectively. The former is rigorously derived from our simulations, while the latter assumes a laminar axisymmetric viscous disk with kinematic viscosity (\ref{eq:viscosity}). }
    \label{fig:cos_accretion_2}
\end{figure}

\begin{figure*}
\centering 
	\includegraphics[width=\textwidth]{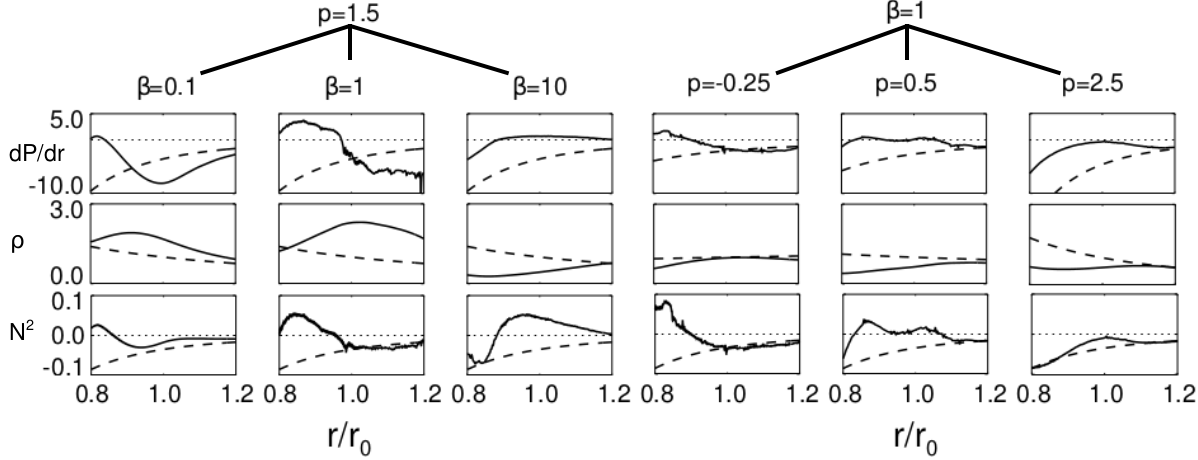}
    \caption{Effect of the nonlinear COS on the radial disk structure. From top to bottom we show initial (dashed curves) and final (after 1,000 orbits, solid curves) profiles of the radial pressure gradient (the vertically and azimuthally averaged pressure), the density and the squared radial buoyancy frequency. The left (right) panels compare different cooling times $\beta$ (initial density slopes $p$).}
    \label{fig:disc_evol}
\end{figure*}

\subsubsection{Vortex migration and spiral density waves}\label{sec:spirals}

 \begin{figure*}
 \centering 
 	\includegraphics[width=\textwidth]{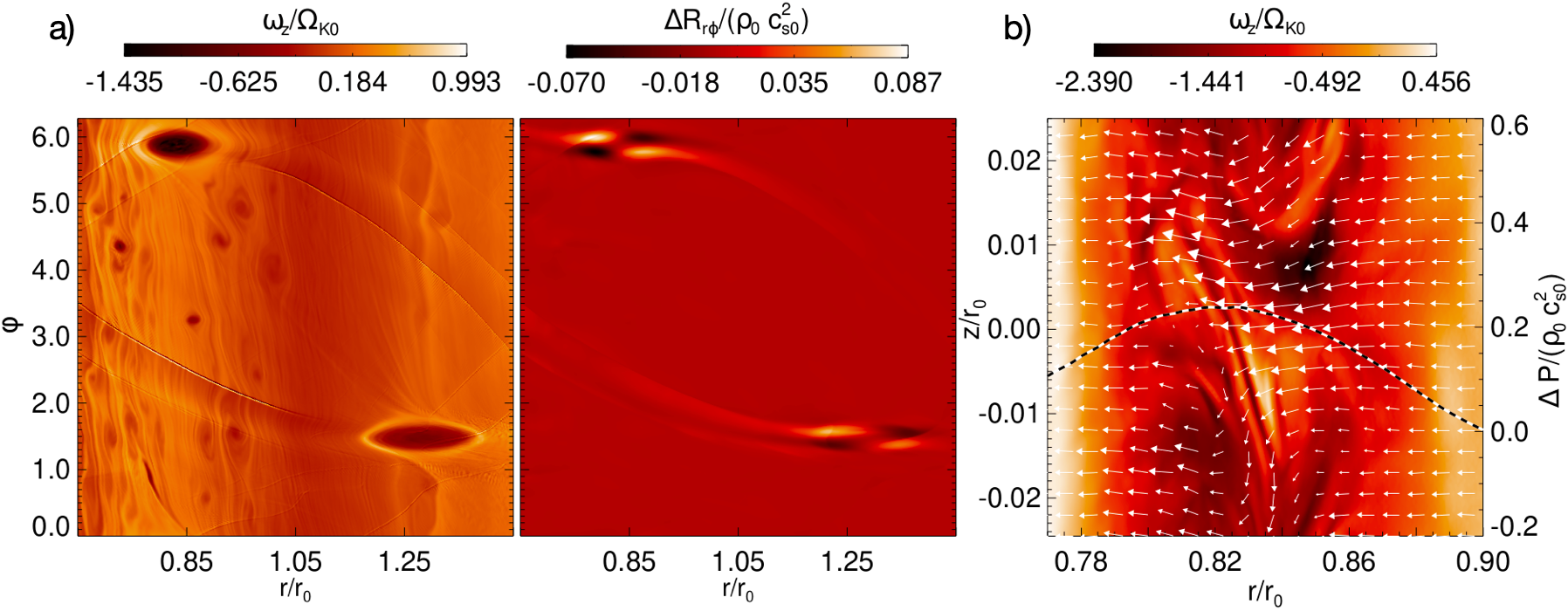}
     \caption{Illustration of the excitation of spiral density waves by strong vortices induced by the COS in the simulation with $p=2.5$. \textbf{a)}: Vertically averaged vertical component of vorticity defined in (\ref{eq:vort}), as well as the quantity $\Delta \mathcal{R}_{r\varphi}$ defined in (\ref{eq:dreynolds}). \textbf{b)}: Vertical structure of the upper left vortex in the left frames. The arrows indicate velocities in the $rz$-plane and the dashed curve denotes the pressure deviation $\Delta P$, showing the pressure maximum associated with the vortex.}
     \label{fig:spirals}
 \end{figure*}

Angular momentum transport in our 3D simulation is presumably realized through the propagation and (shock-)damping of spiral density waves, excited by sufficiently strong vortices (see \citealt{paardekooper2010} for an explanation for the excitation mechanism). This is illustrated in Figure \ref{fig:spirals}, where two strong vortices are present in the computational domain, both of which excite spiral waves. The left panel shows the vertically averaged vertical component of the vorticity
\begin{equation}\label{eq:vort}
    \boldsymbol{\omega} = \nabla \times \delta \boldsymbol{v}, 
\end{equation}
where $\delta \boldsymbol{v}$ is the velocity on top of the equilibrium velocity (\ref{eq:v_eq}).
Furthermore, we consider the quantity (cf. \S \ref{sec:diagnostics})
\begin{equation}\label{eq:dreynolds}
   \Delta \mathcal{R}_{r\varphi}  \equiv  \Delta \left(\rho v_{r}\right) \Delta v_{\varphi} ,
\end{equation}
representing the Reynolds stress induced by non-axisymmetric structures.
  This quantity is displayed in the middle panel of Figure \ref{fig:spirals}, showing that radial angular momentum transport is entirely due to vortices. 
Moreover, the plot in the right panel shows a vertical cut through the upper left vortex, illustrating that strong COS-vortices can possess a significant vertical flow structure.

\begin{figure*}
 \centering 
 	\includegraphics[width=0.65\textwidth]{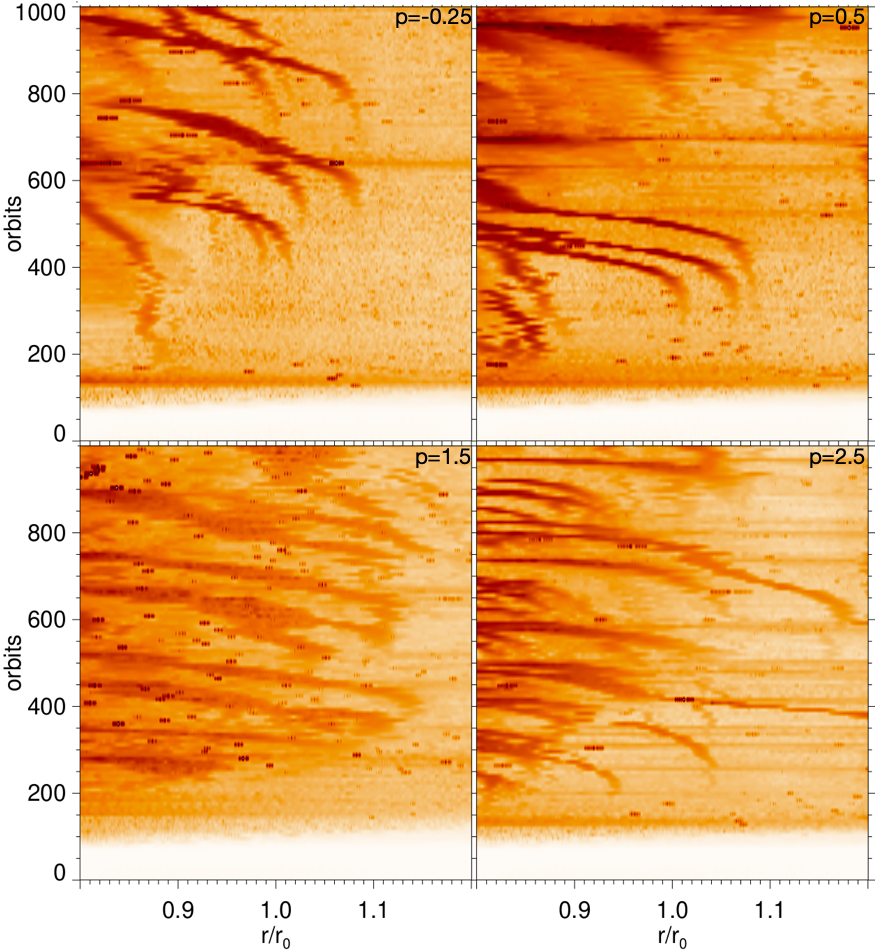}
     \caption{Illustration of the formation and subsequent migration of large scale vortices in 3D simulations with different initial density slopes $p$. The plots show contours of the quantity $\text{min}[\langle \omega_{z}\rangle_{z} - \langle  \omega_{z} \rangle_{z \varphi}]$ (we plot its square root for improved visibility). The dark small-scale features, notably seen in the lower left panel, result from spiral wave shocks which survived our applied smoothing algorithm. }
     \label{fig:migrate}
 \end{figure*}

\begin{figure*}
 \centering 
 	\includegraphics[width=0.9 \textwidth]{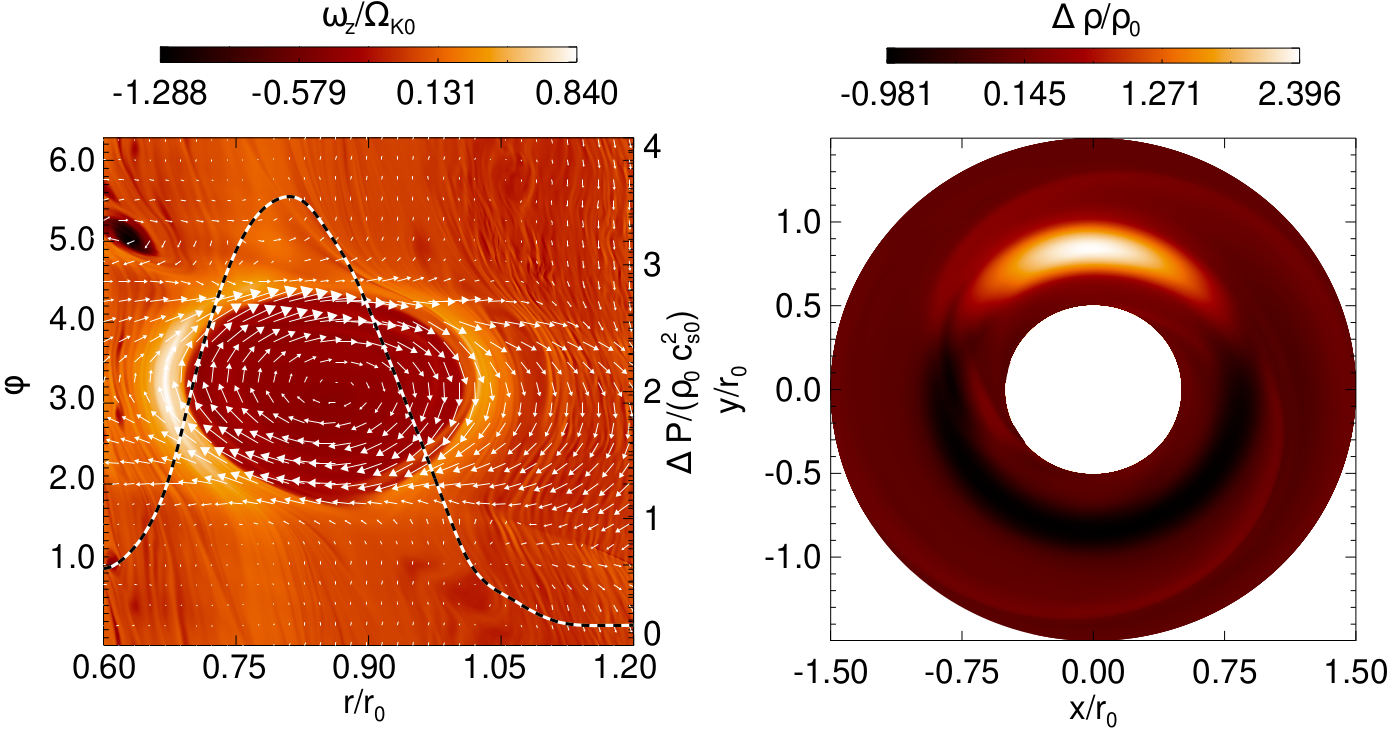}
     \caption{Appearance of a large scale (radial width $\approx 3 H$) long-lived vortex with aspect ratio $\chi\approx 7$ in the 3D simulation adopting a short cooling time $\beta=0.1$ (after $\approx 1,300$ orbits). This vortex exhibits practically no vertical structure and vertical flow field (not shown), in contrast to the vortex shown in Figure \ref{fig:spirals}. The left panel shows the vertically averaged $z$-component of vorticity (\ref{eq:vort}). The arrows illustrate the retrograde flow field associated with the vortex. The over-plotted dashed curve is the scaled pressure deviation $\Delta P$ (at $z=0$), clearly showing a pressure maximum associated with the vortex core. The right hand panel shows a polar plot of the scaled density deviation $\Delta \rho $
     . Spiral waves excited by the vortex are clearly visible in the right panel.}
     \label{fig:large_vortex}
 \end{figure*}

The positive radial flux of angular momentum induced by the vortices leads to radial inward migration of the latter \citep{paardekooper2010}. This is illustrated in Figure \ref{fig:migrate}.  Here, we follow \citet{manger2020} and display the quantity $\text{min}[\langle \omega_{z}\rangle_{z} - \langle  \omega_{z} \rangle_{z \varphi}]$,   
 where we additionally apply a smoothing filter to this quantity to suppress the visibility of sharp features related to spiral waves, and to increase the visibility of larger vortices. The dark diagonal streaks in the figure represent vortices, which undergo rapid inward migration with typical speeds of $0.01H$ per orbit. The faint horizontal bands visible in some panels are spiral waves that survived our smoothing procedure. 
 It appears that vortices are more numerous in our simulations with larger $p \geq 1.5$. Also, in general they more rapidly grow strong in the same simulations, which is reflected in the earlier inward migration after their formation. 
This is consistent with the values of $\alpha_{r}$ and $\dot{M}$ in figure \ref{fig:3D_compare_pslop}.
 Furthermore, merging of vortices can be seen in many cases.
In addition, we find that the COS can, in principle, generate large-scale, long-lived vortices that may be observable in real PPDs based on their size. An example vortex with a radial width $\approx 3 H$ is presented in Figure \ref{fig:large_vortex}. 

Finally, we note that the somewhat unexpected behavior of the simulation with the longer cooling time $\beta=10$ (for which we expect the COS to develop significantly later than in the case with $\beta=1$) could be related to the fact that in the nonlinear state, sub-critical baroclinic vortex amplification \citep{peterson07a,peterson07b,lesur2010} should be active within the vortices in our simulations, and the fact that the latter typically rotate at a much slower rate than the local disk rotation rate, with a turnover frequency $\sim\OmK/\chi$, with the vortex aspect ratio $\chi$, denoting the ratio of the vortex semi-major and semi-minor axes.
 Therefore, the larger the vortex aspect ratio, the longer the rotation time, and as such the longer the optimal cooling time for baroclinic amplification \citep{raettig2013}. Indeed, we find the occurrence of strong and large vortices in the early stages of this simulation before 400 orbits (not shown), which produce spiral density waves which undergo strong shocking. The gradual decay of $\alpha_{r}$, $\dot{M}$ and $\alpha_{z}$ for the simulation with $\beta=10$ after 400 orbits is likely related to the strong modification of the radial disk structure. Figure \ref{fig:disc_evol} shows that $N^2>0$ almost everywhere within the diagnostic domain, such that the COS, and thus the production of new vortices, should be strongly suppressed. Nevertheless, a detailed understanding of the effect of the cooling time scale on COS vortices is beyond the scope of this paper.

\subsubsection{Elliptic instability of vortices}\label{sec:elliptic}

Our simulation results suggest that vortices with a sufficiently small aspect ratio $\chi$ are subjected to the elliptic instability \citep{lesur2009}, which should reduce their lifetimes. 
To illustrate this, we compute the vertically averaged $z$-component of the vorticity (\ref{eq:vort})
and relate it to the vortex aspect ratio via \citep{lesur2010}
\begin{equation}\label{eq:elliptic}
   \text{min}\left[ \langle \omega_{z} \rangle_{z} \right] \equiv -C \frac{3}{2 \chi} \left(\frac{\chi+1}{\chi-1}\right),
\end{equation}
which is valid for a Kida vortex \citep{kida1981} if the constant dimensionless parameter $C=1$. 
Note that, strictly, this simplified correspondence assumes that the minimum vorticity in our simulations occurs within the core of a dominant (or representative) vortex at each time during the simulation. This is a strongly idealized picture for two reasons. For one, several vortices are present in the simulation domain during the nonlinear COS stage. Furthermore, these vortices in general excite spiral waves. The latter form shocks in many cases, which can dominate the minimum and maximum values of the vorticity. 
To determine the proportionality factor $C$, we measure the aspect ratio of a strong and resolved vortex in each of our simulations, using at least three snapshots, and relate it to $\text{min}\left[ \langle \omega_{z} \rangle_{z} \right]$ via Eq. (\ref{eq:elliptic}). To suppress the influence of spiral waves, we apply the same smoothing procedure as in Figure \ref{fig:migrate} above.
We find $C=3.0,2.8,2.4,1.8$ for density slopes $p=-0.25,0.5,1.5,2.5$, respectively. 

For illustration Figure \ref{fig:elliptic} shows the results for the cases $p=0.5$ and $p=2.5$. In this figure, the solid curves represent $\chi$, whereas the dashed curves show the quantity $\text{max}\left[v_{z}\right]-\text{min}\left[ v_{z} \right]$, which should be an indicator for the onset of the elliptic instability, which is expected to attack vortices with $\chi\lesssim 4$ \citep{lesur2010}. Indeed, in many cases where the computed values of the \emph{effective} vortex aspect ratio drop below this value in our simulations, we do find notable increases (decreases) in the maximum (minimum vertical velocity). 
Note that the correspondence mediated by Eq. (\ref{eq:elliptic}) is in some cases compromised by the presence of spiral wave shocks that survive the above-mentioned smoothing procedure and which artificially reduce our computed $\chi$-values.
The insert figure in each panel shows an example of a vortex with $\chi<4$,  which contains fine-scale structures within its core, accompanied by strong vertical velocity fluctuations, indicative of the elliptic instability.

\section{Discussion}\label{sec:discussion}

\subsection{Summary}\label{sec:summary}

We studied the convective overstability near the mid-plane of radially global protoplanetary discs. We first conducted a linear stability analysis. Generally, we find that linear COS modes can only exist radially inward of their corresponding Lindblad resonance. In particular, the fastest growing modes are those whose Lindblad resonances lie within the computational domain, such that their growth rate increases with decreasing resonance radius. Since the modes are not resonantly excited, Lindblad resonances merely serve as turning points in the present context.

The radially global, nonlinear saturation state of the COS in axisymmetric simulations is similar to that in local incompressible shearing box simulations (\hyperlinkcite{teed2021}{TL21}). We find that the flow is dominated by intermittent or persistent zonal flows and elevator flows, for sufficiently large values of pseudo-Richardson number $R$. However, differences occur. 
In \S \ref{sec:smooth_2d} we found that for a viscosity $\alpha \gtrsim 10^{-5}$ the COS is entirely suppressed in our simulations. For $Re=10^5$ we find $R_{\text{crit}}\approx 0.1 $ via Eq. (\ref{eq:ncrit}), which is only marginally largr than the largest values of $R$ (i.e. close to he inner radial domain boundary) for our fiducial parameters. However, based on the results of \hyperlinkcite{teed2021}{TL21} we would still expect the COS to generate some hydrodynamic activity in form of a (disordered) nonlinear wave state in this simulation. 
 The reason for this discrepancy is unclear, but could be related to the (inevitably) different boundary conditions applied in our simulations, as compared to \hyperlinkcite{teed2021}{TL21}.
Nevertheless, the close resemblance of the results for sufficiently large values of $R$ indicates that the nonlinear saturation process of the axisymmetric COS in this parameter regime, which is the most relevant one to planetesimal formation in PPDs, is well captured by a radially local, incompressible model. 

On the other hand, the radially global nature of the COS is not captured in local shearing box models. This is expected to be relevant in more realistic, non-smooth disc setups containing density features such as pressure bumps or gaps. Here, we find, in agreement with our linear calculations, that the COS may be locally excited but that its influence penetrates stable regions radially inwards. Moreover, in simulations applying radial damping zones, we find that the COS can be suppressed entirely (for at least 10,000 orbital periods), which is not expected if it were a radially local instability.

The nonlinear evolution revealed by our 3D simulations appears to be rather complex, and is - in contrast to axisymmetric simulations - no longer governed only by the two quantities $Re$ and $R$ (\S \ref{sec:prelim}). If the COS is sufficiently strong, vortex formation ensues, which can result in significant radial mass and angular momentum transport via the excitation of spiral density waves. This transport modifies the radial disc structure, including forming pressure bumps. Moreover, our results show that the (initial) radial disc structure (described by the two slopes $p$ and $q$) affects the intensity of vortices and spiral waves, and therefore also the magnitude of the ensuing radial mass and angular momentum transport. This is another aspect that local shearing box simulations cannot capture.
Notably, the COS can induce \emph{outward} radial mass transport, i.e., decretion, which is found to be the case in most of our 3D simulations. 
In contrast, in simulations of VSI-active discs, one typically finds an inward mass flux \citep{manger2018,lehmann2022}. Note, though, that such simulations are vertically stratified, unlike those considered here. 

 \begin{figure*}
 \centering 
 	\includegraphics[width=\textwidth]{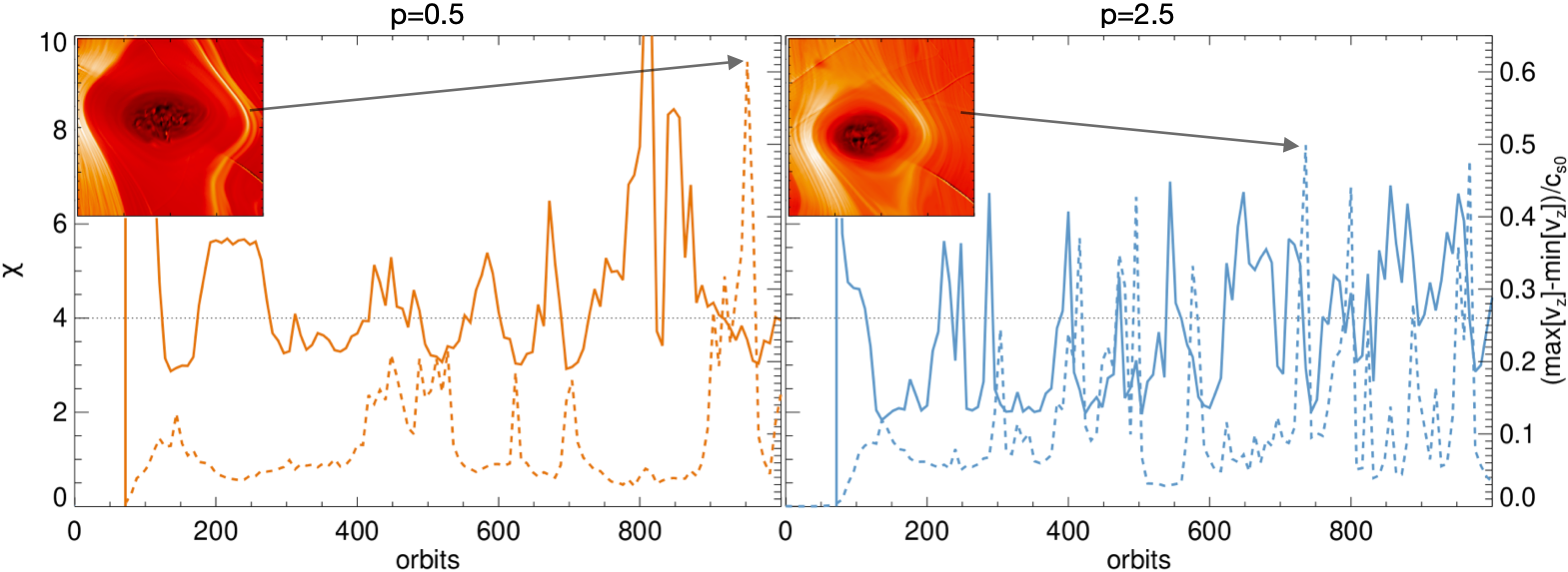}
     \caption{Illustration of the elliptic instability in vortices resulting from the COS. The solid curves show the effective aspect ratio $\chi$ of vortices (\ref{eq:elliptic}) occurring in 3D simulations with different initial density slopes $p$. The dashed curves represent the maximum vertical velocity. The insert figures show examples of vortices where the elliptic instability is suspected to be active. In both cases, the displayed radial region is $0.7<r/r_0 < 1.1$ and the azimuthal extent $\Delta \varphi =2$. Note that vortices around 300 orbits in the simulation with $p=0.5$  are tiny, such that the elliptic instability likely cannot be resolved, explaining the absence of notable increases of $v_{z}$.}
     \label{fig:elliptic}
 \end{figure*}

Our 3D simulations reveal the emergence of large-scale, long-lived vortices. These generally migrate radially inward with rates $\sim 0.01H$ per orbit, similar to vortices generated by the VSI \citep{manger2020,lehmann2022}. 
We also find that strong vortices with small aspect ratios are subject to elliptic instability, which affects their lifetimes.
Moreover, intense vortices, such as the one in Figure \ref{fig:spirals}, can contain significant vertical structure in their flow field, whereas sufficiently weak vortices, such as the one in Figure \ref{fig:large_vortex}, lack any notable vertical structure. 
 These differences are expected to affect the concentration of dust by such vortices. We also find that the COS can, in principle, spawn long-lived vortices (Figure \ref{fig:large_vortex}) that should be large enough to be observable by ALMA and could, therefore, be a possible explanation for observed asymmetries in PPDs.

\subsection{Implications for dust substructures}

Small dust grains tightly coupled to the gas are expected to possess a velocity
\begin{equation}\label{eq:tva}
    \boldsymbol{v}_\mathrm{d} = \boldsymbol{v} + \tau_\mathrm{s} \frac{\nabla P}{\rho_\mathrm{tot}},
\end{equation}
where $\tau_\mathrm{s}$ denotes the particle stopping time and $\rho_\mathrm{tot}$ is total density of the dust-plus-gas mixture \citep{youdin2005}. 
Most of our 3D simulations yield at least one radial location where the radial pressure gradient vanishes (Figure \ref{fig:disc_evol}). For sufficiently small $\left|\boldsymbol{v}\right|$, dust grains respond to the local pressure gradient and are thus expected to drift radially into these regions and, if the pressure bump remains axisymmetric, produce dust rings. Dust rings may also result from `traffic jams' produced from differential dust drift owing to a non-uniform radial pressure gradient, which is almost always the outcome of the COS. 

On the other hand, our 3D simulations also reveal that the nonlinear COS gives rise to an averaged (turbulent) radial gas velocity (cf. Figure \ref{fig:cos_accretion_2}), which, as explained above, in most cases results in mass decretion. Whether this velocity is large enough to overcome pressure gradients and affect the dust concentration in the pressure bumps described remains to be seen in future work.

 Finally,  the COS readily produces vortices, which can trap dust radially and azimuthally \citep{lyra2013,raettig2015,raettig2021}. The COS may, therefore, explain dust rings and asymmetries observed in PPDs \citep{andrews18,long18,vdmarel2021}. However, dust trapping cannot be too significant; otherwise, planetesimal formation would ensue, which is not observable.

\subsection{Non-local particle stirring}

 In the more probable scenario where $N^2<0$ is only realized near particular radii, our results indicate that the COS can still drive activity interior to such regions due to its radially global character. This provides a background turbulence to stir up dust grains. From the vertical Reynolds stress $\text{rms}(\alpha_{z})$ (Figures \ref{fig:3D_compare_beta}-\ref{fig:3D_compare_pslop}), we can estimate the expected height of a passive dust layer $H_\mathrm{d}$ to be (e.g. \citealt{lin2021})
\begin{equation} 
\frac{H_{d}}{H}  \approx   \sqrt{\frac{\text{rms}(\alpha_{z})}{St}}  \approx 10^{-3}St^{-1/2}  .
\end{equation}
with the particle Stokes number $St= \tau_\text{s} \OmK$, where it is assumed that $St\gg \alpha_z$.

Considering a standard Minimum Mass Solar Nebula \citep{chiang2010}, pebbles with an internal density of $1\,\text{g} \text{cm}^{-3}$ and a size of $3 \text{mm}$ would have $St\sim 10^{-3}$ at $1 \text{AU}$, which are then predicted to possess a scale height of $H_\mathrm{d} \approx 0.03H$. If the vertically-integrated metallicity $Z = 0.01$, then the resulting midplane dust-to-gas ratio $\epsilon\sim 0.3$ would be insufficient for the streaming instability to drive strong clumping and hence planetesimal formation should not proceed \citep{johansen2007}. Furthermore, turbulent stirring in of itself stabilizes the streaming instability \citep{umurhan2020,chen2020}. We can, therefore, expect a finite-sized pebble layer to be maintained radially inward from where the COS is locally excited.

\subsection{Caveats and outlook}

In this study, we applied a simple optically thin gas cooling law with a single parameter $\beta$. However, recent studies indicate that cooling times favorable for the COS most likely imply that gas cooling is in the optically thick regime close to the disk mid-plane.
Future simulations should, therefore, consider more realistic gas cooling prescriptions, incorporating thermal diffusion. 
Furthermore, as our simulations are vertically unstratified, they omit the possible presence of the VSI, which may occur for the parameter regimes considered here, if vertical stratification and more realistic cooling were accounted for. 
Fully global models are required to study the possible interplay of the COS and the VSI in a realistic manner.

Our example simulation in which the COS results from the presence of a pressure bump shows the formation of weak non-axisymmetries and vortices only after ten thousands of orbits. It is, therefore, speculative whether the required time scales are relevant to planetesimal formation in PPDs.  It further remains speculative if the resulting vortices can efficiently concentrate dust.
Also, more detailed observational constraints on the radial density and temperature structure of PPDs are required to determine if the COS is potentially relevant to planetesimal formation and whether or not it is only expected to occur near certain density structures such as pressure bumps.

Due to limitations in computational resources, our results on the radial mass transport in 3D simulations are not converged with respect to spatial resolution (see Appendix \ref{app:convergence}), and, perhaps more importantly, vertical domain size (not shown). The latter implies that a quantitative study of the radial mass transport due to the COS formally requires a vertically stratified disk model. Such simulations will also need to adopt even higher grid resolutions than in this work, which poses a great computational challenge.

We note that we do not fully understand the reasons behind the exceptionally large radial mass flux occurring in our fiducial 3D simulation with $p=1.5$, as compared to the other simulations. We find (not shown) that in the former simulation the $\chi$-values computed via (\ref{eq:elliptic}) seem to hover on the threshold for marginal stability (i.e. $\chi\approx 4$), unlike the other simulations (cf. Figure \ref{fig:elliptic}), where $\chi$ drops well below the threshold. Perhaps the vortices in the simulation with $p=1.5$ are marginally unstable to the elliptic instability, and might therefore survive longer as those in the other simulations. Moreover, we find a relatively large pileup of mass in the diagnostic domain in this simulation, compared to the other simulations (see Figure \ref{fig:cos_accretion_1}). This explains to some extent the large mass flux.
However, it remains unclear how this behaviour is connected to the equilibrium density slope $p$. We speculate that the radial vortencity profile $\rho/\omega_{z}$, which yields a constant in the special case $p=1.5$ plays a role in the evolution of vortices in our simulations. We therefore suggest that future work should study the evolution of vortices in radially global disk models in a more systematic manner. 

Note also that our discussion on the elliptic instability in \S \ref{sec:elliptic} is not rigorous.
As mentioned earlier, the correspondence mediated by Eq. (\ref{eq:elliptic}) is in some cases compromised by the presence of spiral wave shocks that survive our smoothing procedure and which artificially reduce our computed $\chi$-values. 
Moreover, the measurements of vortex aspect ratios are done by hand and are, therefore, subject to significant potential inaccuracy. Furthermore, vortices in our simulations are not necessarily Kida vortices, which are defined by a constant patch of relative vorticity. Nevertheless, our results strongly suggest that the elliptic instability occurs in the cores of COS-vortices of sufficiently small aspect ratio.

In a follow-up study, we will investigate the impact of the nonlinear COS on dust concentration in PPDs. In particular, we will clarify if COS vortices in radially global PPD models can result in dust-to-gas density ratios large enough to trigger vigorous streaming instability.

\begin{acknowledgments}

We thank an anonymous reviewer for useful comments.
This work is supported by the National Science and Technology
Council (grants 112-2112-M-001-064-, 
112-2124-M-002-003-, 113-2124-M-002-003-) and an Academia
Sinica Career Development Award (AS-CDA-110-M06). Simulations were performed on the \emph{Kawas} cluster at ASIAA and the \emph{Taiwania-2} cluster at the National Center for High-performance Computing (NCHC). We thank NCHC for
providing computational and storage resources. 

\end{acknowledgments}

\newpage

\appendix

\section{COS in local compressible model}\label{app:compress}

For completion we provide the linearised equations in a local compressible model (cf. \citealt{lyra2014}). These are obtained from the radially global equations (\ref{eq:linQ})---(\ref{eq:linW}) by replacing $\partial_r \to i k_r$ and using (\ref{eq:rho_equ})-(\ref{eq:T_equ}). The resulting equations constitute a linear eigenvalue problem
\begin{align}\label{eq:eigenproblem}
   \boldsymbol{M} \boldsymbol{b} = \omega \boldsymbol{b},
\end{align}
with
\[\boldsymbol{M}=
\begin{bmatrix}
  -i \omega & cs^2 \left(1+ k_r -p\right) & 0 & -i k_z c_s^2 & 0 \\
    \frac{2\eta}{c_s^2} & -i \omega & -2 & 0 & i k_r -p \\
    0 & 2 + r\partial_r \Omega & -i \omega & 0 & 0 \\
    0 & 0 & 0 & -i \omega & i k_z \\
    -\frac{1}{\beta} & -2 \eta +c_s^2 \gamma \left(1+i k_r\right) & 0 & i c_s^2 \gamma k_z & -i \omega + \frac{1}{\beta} \\
\end{bmatrix},
\]
eigenvector $\boldsymbol{b}=\left\{Q, \delta v_{x}, \delta v_y, \delta v_z, W \right\}^T$ and eigenvalue $\omega$. 
The linear growth rates of inertial waves resulting from this set of equations are largely similar to those resulting from an incompressible Boussinesq model \citep{latter2016,lehmann2023} and will not be presented here for brevity. 

\section{Linear problem including viscosity}\label{app:visc}

The linearised radially global momentum equations including a \emph{constant}\footnote{this is chosen so as to facilitate comparison with our simulations.} kinematic shear viscosity $\nu$ are given by
\begin{equation}\label{eq:linvr_visc}
\begin{split}
   i \omega \delta v_{r} & = -2 \Omega \delta v_{\varphi} - \frac{\partial_{r} P}{\rho c_{s}^2} Q + \partial_{r} W + \partial_{r} \ln \rho W\\
   \quad & \nu\big[\frac{4}{3 r}\delta v_{r}^{'} + \frac{4}{3} \partial_{r} \delta v_{r}^{'} -(\frac{4}{3 r^2} +k_{z}^2) \delta v_{r} + \frac{1}{3}i k_{z} \delta v_{z}^{'}\\
   \quad & +  \partial_{r}  \ln \rho\left(\frac{4}{3} \delta v_{r}^{'} -\frac{2}{3 r}\delta v_{r} -\frac{2}{3} i k_{z} \delta v_{z} \right) \big]
\end{split}
\end{equation}
\begin{equation}\label{eq:linvphi_visc}
\begin{split}
   i \omega \delta v_{\varphi} & = \left(2\Omega + r \partial_{r} \Omega \right) \delta v_{r} \\
   \quad & + \nu\big[\frac{1}{r}\delta v_{\varphi}^{'} + \partial_{r}\delta v_{\varphi}^{'}-(\frac{1}{r^2} + k_{z}^2)\delta v_{\varphi}\\
   \quad & + \partial_{r} \ln \rho \left(\partial_{r} \delta v_{\varphi} -\frac{1}{r}\delta v_{\varphi}\right)\\
   \quad & - \left(\partial_{r} \ln \rho \frac{\delta \rho}{\rho} - \frac{1}{\rho}\partial_{r} \delta \rho \right)r\partial_{r}\Omega\big], 
   \end{split}
\end{equation}
\begin{equation}\label{eq:linvz_visc}
\begin{split}
    i \omega \delta v_{z} & = i k_{z} W + \nu \big[-\frac{4}{3}k_{z}^2 \delta v_{z} + \partial_{r} \delta v_{z}^{'} + \frac{1}{3}i k_{z} \delta v_{r}^{'} \\
    \quad & + \frac{\delta v_{z}^{'}}{r} + \frac{i k_{z}}{3 r} \delta v_{r} +\partial_{r}\ln \rho \left(\partial_{r} \delta v_{z} + i k_{z} \delta v_{r}\right)\big],
       \end{split}
\end{equation}
where a prime ($'$) indicates a radial derivative, and where we neglected the equilibrium radial viscous inward drift:
\begin{equation}\label{eq:vr_visc}
    v_{r} =  \frac{\nu}{2\Omega + r \partial_{r}\Omega}\left([3-p] \partial_{r}\Omega + r \partial_{r}^2 \Omega \right),
\end{equation}
which follows from the azimuthal component of Eq. (\ref{eq:contvg}) at equilibrium.
In order to solve the viscous problem (\ref{eq:linQ}), (\ref{eq:linW}),  (\ref{eq:linvr_visc})---(\ref{eq:linvz_visc}) using the pseudo-spectral method (\S \ref{sec:lin_num}) we now need to specify five additional boundary conditions.
We find that
\begin{align}
    \delta v_{r}^{'}(r_{\text{min}})&=0,\\
    \delta v_{y}^{'}(r_{\text{min}})&=\delta v_{y}^{'}(r_{\text{max}})=0,\\
       \delta v_{z}^{''}(r_{\text{min}})&=0,\\
           W(r_{\text{max}})&=0,
\end{align}
is a suitable set of boundary conditions.

On the other hand, to apply the shooting method we now consider the set of first order ODEs:
\begin{equation}
    \begin{split}
        \partial_{r} U & = i \omega \delta v_{r} + 2 \Omega \delta v_{\varphi} + \frac{\partial_{r}P}{\rho c_{s}^2}Q -\partial_{r} \ln \rho W \\
        \quad & -\left(\frac{1}{r}+ \partial_{r} \ln\rho \right)\left(U-W\right) \\
        \quad & +\nu\bigg[ \left(\frac{4}{3} \frac{1}{r^2} +\frac{2}{3}\frac{\partial_{r}\ln\rho}{r} +k_{z}^2\right) \delta v_{r} - \frac{1}{3}i k_{z} \delta v_{z}^{'} \\
        \quad & + \frac{2}{3}\partial_{r} \ln \rho i k_{z}\delta v_{z}\bigg] ,
    \end{split}
\end{equation}
\vspace{-0.2cm}
\begin{equation}
    \begin{split}
        \partial_{r} \delta v_{\varphi}^{'} & = \frac{1}{\nu}\left[i\omega \delta v_{\varphi} -\left(2\Omega + r \partial_{r} \Omega\right)\delta v_{r}\right]\\
        \quad & -\frac{Q}{c_{s}^2}\partial_{r}\Omega - \left(\frac{1}{r} + \partial_{r} \ln \rho\right)\delta v_{\varphi}^{'} \\
        \quad & + \left(\frac{1}{r^2} + k_{z}^2 + \frac{\partial_{r}\ln\rho}{r}\right)\delta v_{\varphi},
    \end{split}
\end{equation}
\begin{equation}
    \begin{split}
        \partial_{r} \delta v_{z}^{'} & = \frac{1}{\nu}\left[ i \omega \delta v_{z} -\frac{3}{4}i k_{z}W -\frac{1}{4}i k_{z} U\right]\\
        \quad & + \frac{4}{3}k_{z}^2 \delta v_{z} -\frac{1}{r}\delta v_{z}^{'}-\frac{i k_{z}}{3 r}\delta v_{r}\\
        \quad & -\partial_{r}\ln\rho\left(\delta v_{z}^{'} + i k_{z} \delta v_{r} \right),
    \end{split}
\end{equation}
\begin{align}
    \partial_{r}\delta v_{r} &= \frac{3}{4\nu}\left(U-W\right),\label{eq:linU}\\
    \partial_{r}\delta v_{\varphi} &= v_{\varphi}^{'},\\
    \partial_{r}\delta v_{z} &= v_{z}^{'},
\end{align}
where we defined the new variable $U$ via Eq. (\ref{eq:linU}),
and where 
\begin{equation}
        Q = \frac{A}{B}W +\frac{\gamma c_{s}^2 \partial_{r}\ln \rho-\frac{\partial_{r}P}{\rho}}{B}\delta v_{r},
\end{equation}
following from (\ref{eq:linW}), as well as
\begin{equation}
    \begin{split}
        W & = \left(1+ \frac{4 \nu i \omega A}{3 c_{s}^2 B}\right)^{-1}\frac{4\nu}{3}\bigg[\bigg(-\frac{i\omega}{c_{s}^2}\frac{\gamma c_{s}^2 \partial_{r} \ln\rho -\frac{\partial_{r}P}{\rho}}{B} \\
        \quad & +(\partial_{r}\ln\rho+\frac{1}{r})\bigg)\delta v_{r}  -\left(i k_{z} \delta v_{z} + \frac{3}{4\nu}U\right)\bigg]
    \end{split}
\end{equation}
following from (\ref{eq:linQ}).
Furthermore, we defined the short notations
\begin{align}
    A &\equiv i\omega -\frac{1}{t_{c}},\\
    B &\equiv i\omega\gamma-\frac{1}{t_{c}}.
\end{align}

\section{Numerical convergence of hydrodynamic simulations}\label{app:convergence}

\subsection{2D axisymmetric simulations}
\begin{figure}
\centering 
	\includegraphics[width=0.49 \textwidth]{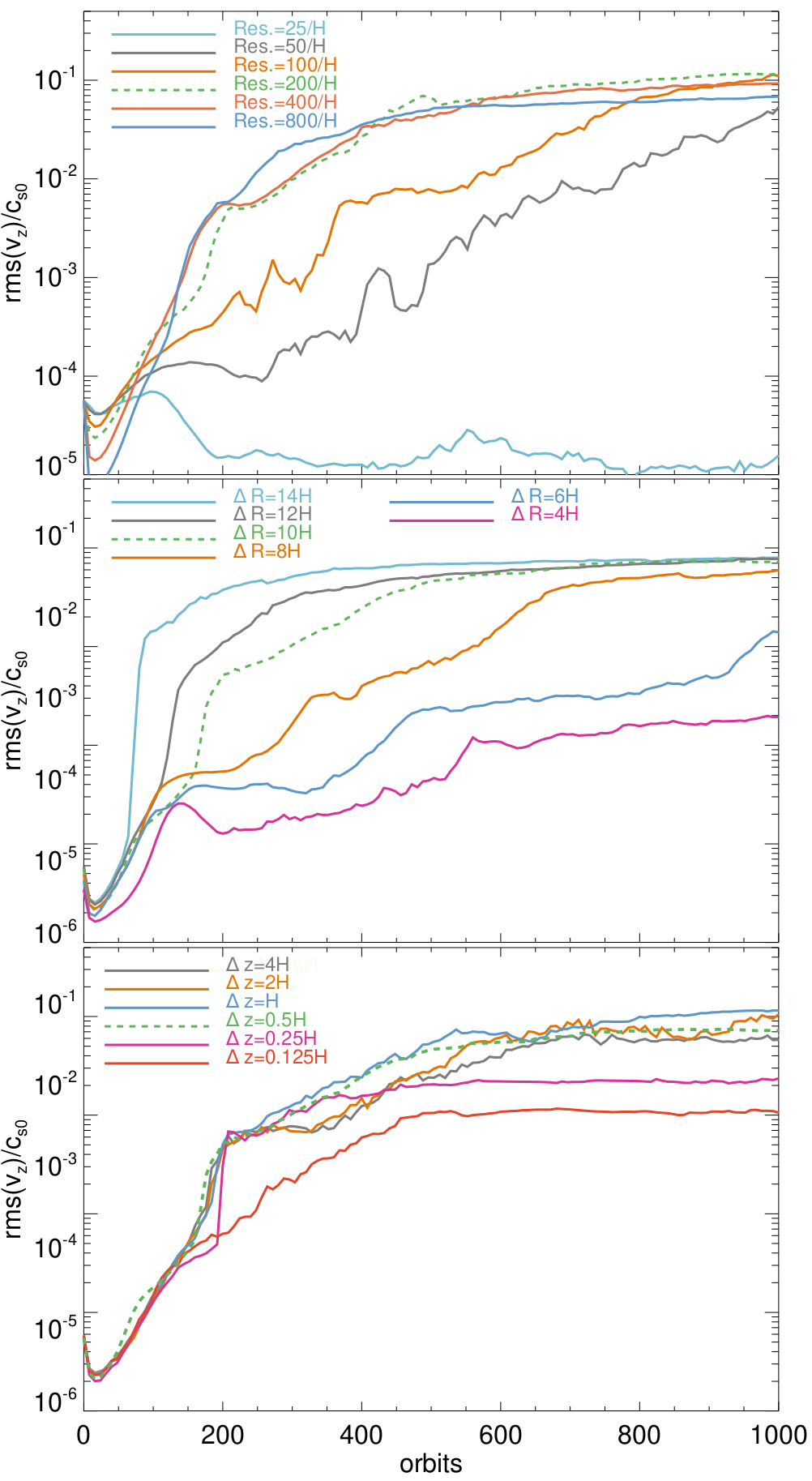}
    \caption{Testing convergence of 2D axisymmetric simulations with respect to resolution (top panel), radial domain size (middle panel) and vertical domain size (bottom panel). 
The simulations in the top panel were conducted using a vertical domain size $\Delta z=H$. Note that all of these simulations adopted wave damping boundaries.}
    \label{fig:cos_var_num}
\end{figure}

Figure \ref{fig:cos_var_num} shows results of 2D simulations with different values of numerical parameters. These are the spatial resolution (upper panel), the radial domain size (middle panel) and the vertical domain size (bottom panel).
The plots in the upper panel demonstrate that our 2D simulations, using a resolution of $200/H$ are converged.
The plots in the middle panel reveal that the saturation level of the COS generally reduces with decreasing radial domain size.
Based on the results of our linear analysis in \S \ref{sec:lin_res_fid} this is expected, and is due to the fact that the maximum value of $R$ defined in (\ref{eq:richardson}) is reduced (the value at the inner domain boundary), as well as our finding that for vertical wavenumbers $k_{z} H_{0} \gtrsim 10$ the fastest growing linear modes (i.e. the modes that can "access" the largest $R$-values) are strongly concentrated towards the inner disk boundary. Thus, with decreasing radial domain size these modes are gradually removed.

We also find a significant drop in the saturation level when decreasing the vertical domain size as $\Delta z =0.5 H_{0}\to0.25 H_{0}$, whereas the corresponding drop is only modest for the reduction $\Delta z = H_{0}\to0.5 H_{0}$. These reductions imply that we increase the minimum vertical wavenumbers $k_{z,\text{min}}=4 \pi \to 8 \pi$ and  $k_{z,\text{min}}=2 \pi \to4 \pi$, respectively.
Considering Figure \ref{fig:disprel_fid} (left panel) these findings are plausible, since the largest growing modes in our simulations possess $10 \lesssim k_{z} H_{0} \lesssim 100$.

\subsection{3D simulations}

\begin{figure}
\centering 
	\includegraphics[width=0.5 \textwidth]{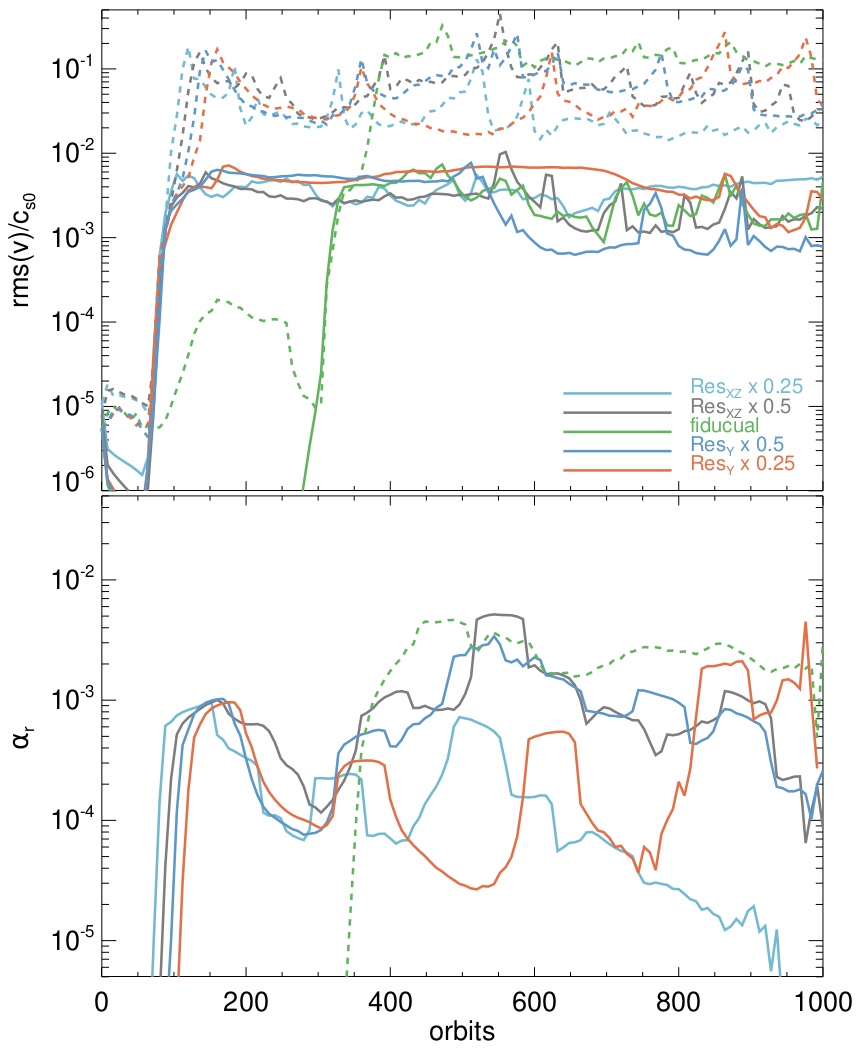}
    \caption{Testing convergence with respect to spatial resolution in 3D simulations. We consider convergence simultaneously in radial and vertical direction (labeled "XZ") as well as azimuthal direction (labeled "Y"). The upper panel shows rms vertical (solid curves) and radial (dashed curves) velocities. The lower panel shows the time evolution of $\alpha_{r}$. The curves in the lower panel have been smoothed for clarity. The dashed curve in the lower panel describes the fiducial simulation.}
    \label{fig:res_3D}
\end{figure}

As pointed out in \S \ref{sec:summary}, the radial angular momentum flux in our 3D simulations is not converged with respect to resolution in all spatial dimensions, which is illustrated in Figure  \ref{fig:res_3D} (lower panel). This can likely be traced back to the formation of spiral shocks (which generate very sharp gradients in the hydrodynamic quantities), which are ultimately responsible for the deposition of angular momentum in the disk.
On the other hand, the plot in the upper panel shows that rms radial (dashed curves) and vertical (solid curves) velocities resulting from the COS are actually converged.
Our chosen resolution of 200/H in both radial and vertical directions proves computationally demanding in 3D global simulations, making it currently impractical to pursue an even higher resolution. Nevertheless, as already noted in \S \ref{sec:summary}, a detailed study of the radial mass flux requires a vertically stratified disk model.

\begin{figure}
\centering 
	\includegraphics[width=0.485 \textwidth]{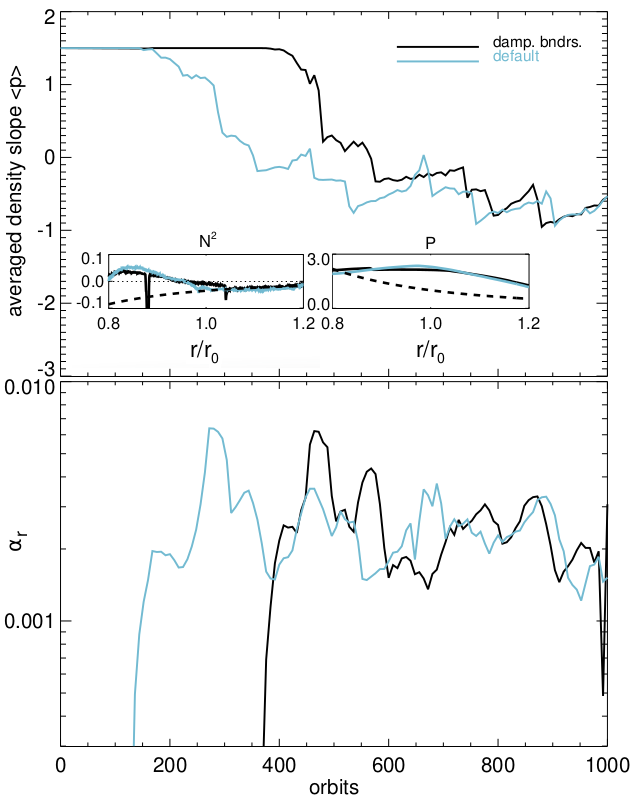}
    \caption{The plots illustrate the robustness of our simulation results with respect to the radial boundary conditions.}
    \label{fig:cos_bndrs}
\end{figure}

Figure \ref{fig:cos_bndrs} illustrates the robustness of our (3D) simulation results with respect to the radial boundary conditions. This is not a straight forward issue in the present study, as the fastest growing COS modes are concentrated towards the inner domain boundary. The plots show the time evolution of the averaged density slope $\partial \ln \rho/ \partial \ln r$ (upper panel) and the $\alpha$-viscosity (lower panel). Compared are fiducial simulations with and without wave damping radial boundary conditions. We find that the application of a damping zone near the boundaries merely results in a delayed nonlinear saturation of the COS, of equal magnitude as in the simulation with default boundaries. The insert plots in the upper panel show the initial and final radial profiles of the squared buoyancy frequency $N^2$ and pressure $P$, in good agreement between the two simulations.

\section{Double-diffusive instability in slowly cooled viscous  disks}\label{sec:double_diffusive}

\subsection{Occurrence in hydrodynamic simulations}

In \S \ref{sec:smooth_2d} we found that rms vertical velocities in 2D simulations do not show the expected behavior of the COS with increasing cooling time. That is, for large cooling times $\beta\gtrsim 100$ they exhibit an unexpected rise.
 We suspect this to be caused by the presence of another instability. In the following we briefly show that a "double-diffusive" instability (not the COS) can operate in the adiabatic limit and in the presence of an unstable entropy gradient, as well as a considerable shear viscosity.  It is double-diffusive in that is relies on a large (viscous) diffusion of angular momentum and small thermal diffusion.
 
\begin{figure*}
 	\includegraphics[width= 0.95\textwidth]{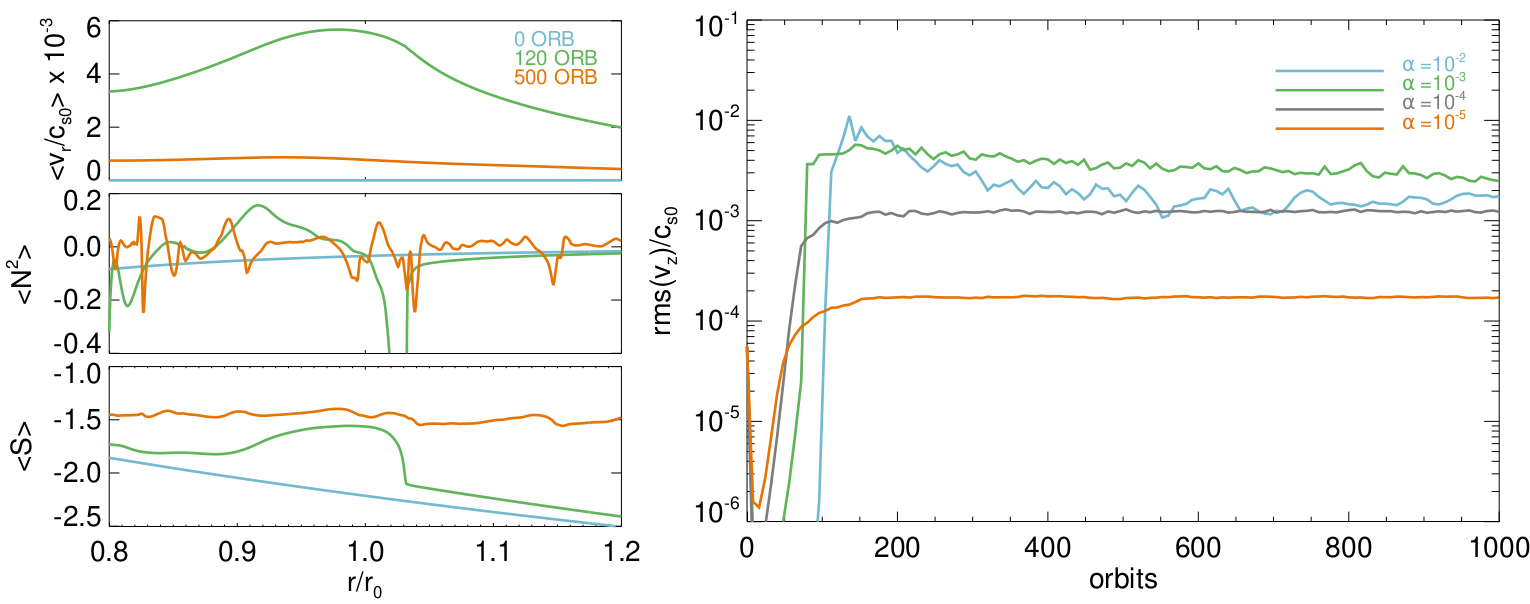}
     \caption{Illustration of the nonlinear saturation of the double diffusive instability powered by viscous momentum diffusion and a negative radial entropy gradient. The right panel shows rms vertical velocities in simulations with increasing kinematic shear viscosity $\alpha$, defined via (\ref{eq:reynolds}). The left panels show the evolution of the vertically averaged radial velocity, squared buoyancy frequency and entropy, in the simulation with $\alpha=10^{-3}$.}
     \label{fig:visc_adiab_instab}
 \end{figure*}

We start by presenting in Figure \ref{fig:visc_adiab_instab} rms vertical velocities (right panel) of adiabatic simulations ($\beta=10^4$) and constant viscosity taking values $\alpha=10^{-5}-10^{-2}$. In these simulations we use $q=2$ and $p=0.5$, which results in $N^2 <0$, and a negligible equilibrium viscous accretion flow according to Eq.(\ref{eq:vr_visc}). We see that for $\alpha\gtrsim 10^{-4}$ the system settles on appreciable rms velocity values. Furthermore, the left panels show from top to bottom radial profiles of the vertically averaged radial velocity, squared buoyancy frequency and entropy, respectively.
As seen, the instability gives rise to a radially varying radial outward velocity, such that (due to the lack of cooling) entropy can be transported outwards, which results in a flattening of the entropy profile. The buoyancy frequency develops strong, stochastic deviations from the ground state profile. However, the averaged value of $N^2$ (over height and radius) saturates to a finite negative value (not shown), explaining the persistence of the instability in the nonlinear regime. Note that the instability similarly develops for different values of $p$. The difference is that the radial viscous accretion velocity (\ref{eq:vr_visc}) adds up to the radial velocity field induced by the instability. Depending on the sign of the former (which depends on $p$) this somewhat speeds up or slows down the flattening of the entropy profile.

\subsection{Local incompressible description}
The existence of such an instability was already predicted by \citet{latter2010b}, 
 who studied a "resistive double-diffusive" instability in magnetized PPDs. The difference is that the version discussed here makes use of the laminar Navier-Stokes viscosity instead of a magnetic field to diffuse angular momentum of fluid parcels, in order to eliminate the stabilising effect of rotation.
 
Let us consider the linearised equations for a gas within the local incompressible Boussinesq model of \citet{lp2017}, as given in \citet{lehmann2023} (their Eqs. (39)-(41)),  
for simplicity presented in the limit\footnote{It can be verified  numerically that the linear instability persists in this limit.} $k_{x}\to 0$ and $\beta \to \infty$ (adiabatic limit), as well as vanishing ground state velocity of the gas:
\begin{align}
   \partial_{t}\delta \rho & = \frac{\rho_{0}}{H_{r}}\delta v_{x} \label{eq:bous_ent},\\
   \partial_{t} \delta v_{x} &= 2 \delta v_{y} -H_{r} N^2 \frac{1}{\rho_{0}}\delta \rho -k_{z}^2 \alpha h_{0}^2\delta v_{x}\label{eq:bous_ux},\\
   \partial_{t} \delta v_{y} & = -\frac{1}{2}\delta v_{x} -k_{z}^2 \alpha h_{0}^2 \delta v_{y},
\end{align}
where the radial entropy stratification is described by the quantity 
\begin{equation}
    \frac{1}{H_{r}}\equiv\frac{1}{\gamma}\frac{\partial S}{\partial r}
\end{equation}
and $N^2$ is given by (\ref{eq:nr2}). Furthermore, $\alpha$ is a dimensionless viscosity parameter.
The above equations are defined in a Cartesian frame $(x,y,z)$ within 
small rectangular patch of the disk, centered on a fiducial point $(r_0,\phi_0 - \Omega_0 t, z=0)$, that rotates around the star at approximately the local Keplerian frequency. The unit vectors $\vec{\mathbf{e}}_x,\, \vec{\mathbf{e}}_y, \, \vec{\mathbf{e}}_z$ in the box corresponding to the radial, azimuthal, and vertical directions in the global disk. 

In the \emph{inviscid} case $\alpha\to 0$, by taking the time derivative of Eq. (\ref{eq:bous_ux}) we find
\begin{equation}
    \partial_{t}^2 \delta v_{x} = -\left(1+N^2\right) \delta v_{x} 
\end{equation}
which yields stable, buoyancy-adjusted epicycles.
However, if we assume that the viscosity is non-zero and consider an initial radial velocity perturbation $\delta v_{x}>0$.
Then, for sufficiently large $k_{z}$ the viscous term in the azimuthal momentum equation prevents the azimuthal velocity $\delta v_{y}$ from becoming large enough to provide a restoring motion via the Coriolis term in the radial momentum equation. 
At the same time, the buoyancy term is found to be positive via (\ref{eq:bous_ent}) and needs to exceed the viscous damping term in (\ref{eq:bous_ux}) in order for $\delta v_{x}$ to grow. The ensuing instability is thus expected to be non-oscillatory, which can easily be confirmed as follows.
The dispersion relation for local modes $\propto      \exp{\left(i k_{x} x + i k_{z} z - i \omega t\right)}$
with $k_{x}=0$ following from above equations reads
\begin{equation}
    \omega^3 +\omega^2 \,2 i k_{z}^2 \alpha h_{0}^2 -\omega\left(1+N^2 + k_{z}^4 \alpha^2 h_{0}^4 \right) -i k_{z}^2 \alpha h_{0}^2 N^2 =0.
\end{equation}
If we assume purely growing modes with $\omega \equiv i \omega_{I}$ we find that all coefficients of the resulting dispersion relation are positive, such that the condition
\begin{equation}
    k_{z}^2 \alpha h_{0}^2 N^2<0
\end{equation}
is a sufficient condition for instability, expressing the necessity of a non-vanishing viscosity \emph{and} an unstable entropy gradient with $N^2<0$.

Nevertheless, even though the linear instability can be understood within the local Boussinesq framework, its nonlinear saturation involves considerable changes of the background disk structure, as shown above. Furthermore, the relevance of this instability to PPDs is questionable, since the considered values of $\alpha$ would require a significant level of turbulence generated by some other source in the first place.

\bibliography{lit}{}

\begin{thebibliography}{}
\expandafter\ifx\csname natexlab\endcsname\relax\def\natexlab#1{#1}\fi
\providecommand{\url}[1]{\href{#1}{#1}}
\providecommand{\dodoi}[1]{doi:~\href{http://doi.org/#1}{\nolinkurl{#1}}}
\providecommand{\doeprint}[1]{\href{http://ascl.net/#1}{\nolinkurl{http://ascl.net/#1}}}
\providecommand{\doarXiv}[1]{\href{https://arxiv.org/abs/#1}{\nolinkurl{https://arxiv.org/abs/#1}}}

\bibitem[{{Andrews} {et~al.}(2009){Andrews}, {Wilner}, {Hughes}, {Qi}, \& {Dullemond}}]{andrews2009}
{Andrews}, S.~M., {Wilner}, D.~J., {Hughes}, A.~M., {Qi}, C., \& {Dullemond}, C.~P. 2009, \apj, 700, 1502, \dodoi{10.1088/0004-637X/700/2/1502}

\bibitem[{{Andrews} {et~al.}(2018){Andrews}, {Huang}, {P{\'e}rez}, {Isella}, {Dullemond}, {Kurtovic}, {Guzm{\'a}n}, {Carpenter}, {Wilner}, {Zhang}, {Zhu}, {Birnstiel}, {Bai}, {Benisty}, {Hughes}, {{\"O}berg}, \& {Ricci}}]{andrews18}
{Andrews}, S.~M., {Huang}, J., {P{\'e}rez}, L.~M., {et~al.} 2018, \apjl, 869, L41, \dodoi{10.3847/2041-8213/aaf741}

\bibitem[{{Armitage}(2011)}]{armitage2011}
{Armitage}, P.~J. 2011, \araa, 49, 195, \dodoi{10.1146/annurev-astro-081710-102521}

\bibitem[{{Bai} \& {Stone}(2010)}]{bai2010}
{Bai}, X.-N., \& {Stone}, J.~M. 2010, \apj, 722, 1437, \dodoi{10.1088/0004-637X/722/2/1437}

\bibitem[{{Balbus} \& {Hawley}(1991)}]{balbus1991}
{Balbus}, S.~A., \& {Hawley}, J.~F. 1991, \apj, 376, 214, \dodoi{10.1086/170270}

\bibitem[{{Barker} \& {Latter}(2015)}]{barker2015}
{Barker}, A.~J., \& {Latter}, H.~N. 2015, \mnras, 450, 21, \dodoi{10.1093/mnras/stv640}

\bibitem[{{Barranco} {et~al.}(2018){Barranco}, {Pei}, \& {Marcus}}]{barranco2018}
{Barranco}, J.~A., {Pei}, S., \& {Marcus}, P.~S. 2018, \apj, 869, 127, \dodoi{10.3847/1538-4357/aaec80}

\bibitem[{{Ben{\'\i}tez-Llambay} {et~al.}(2019){Ben{\'\i}tez-Llambay}, {Krapp}, \& {Pessah}}]{llambay2019}
{Ben{\'\i}tez-Llambay}, P., {Krapp}, L., \& {Pessah}, M.~E. 2019, \apjs, 241, 25, \dodoi{10.3847/1538-4365/ab0a0e}

\bibitem[{Benítez-Llambay \& Masset(2016)}]{fargo3d}
Benítez-Llambay, P., \& Masset, F.~S. 2016, The Astrophysical Journal Supplement Series, 223, 11.
\newblock \url{http://stacks.iop.org/0067-0049/223/i=1/a=11}

\bibitem[{{Blum}(2018)}]{blum2018}
{Blum}, J. 2018, \ssr, 214, 52, \dodoi{10.1007/s11214-018-0486-5}

\bibitem[{Burns {et~al.}(2020)Burns, Vasil, Oishi, Lecoanet, \& Brown}]{burns2020}
Burns, K.~J., Vasil, G.~M., Oishi, J.~S., Lecoanet, D., \& Brown, B.~P. 2020, Phys. Rev. Res., 2, 023068, \dodoi{10.1103/PhysRevResearch.2.023068}

\bibitem[{{Carrera} {et~al.}(2021){Carrera}, {Simon}, {Li}, {Kretke}, \& {Klahr}}]{carrera2021a}
{Carrera}, D., {Simon}, J.~B., {Li}, R., {Kretke}, K.~A., \& {Klahr}, H. 2021, \aj, 161, 96, \dodoi{10.3847/1538-3881/abd4d9}

\bibitem[{{Chen} \& {Lin}(2020)}]{chen2020}
{Chen}, K., \& {Lin}, M.-K. 2020, \apj, 891, 132, \dodoi{10.3847/1538-4357/ab76ca}

\bibitem[{{Chiang} \& {Youdin}(2010)}]{chiang2010}
{Chiang}, E., \& {Youdin}, A.~N. 2010, Annual Review of Earth and Planetary Sciences, 38, 493, \dodoi{10.1146/annurev-earth-040809-152513}

\bibitem[{{Drazkowska} {et~al.}(2022){Drazkowska}, {Bitsch}, {Lambrechts}, {Mulders}, {Harsono}, {Vazan}, {Liu}, {Ormel}, {Kretke}, \& {Morbidelli}}]{drazkowska2022}
{Drazkowska}, J., {Bitsch}, B., {Lambrechts}, M., {et~al.} 2022, arXiv e-prints, arXiv:2203.09759.
\newblock \doarXiv{2203.09759}

\bibitem[{{Fukuhara} {et~al.}(2021){Fukuhara}, {Okuzumi}, \& {Ono}}]{fukuhara2021}
{Fukuhara}, Y., {Okuzumi}, S., \& {Ono}, T. 2021, \apj, 914, 132, \dodoi{10.3847/1538-4357/abfe5c}

\bibitem[{{Gammie}(1996)}]{gammie1996}
{Gammie}, C.~F. 1996, \apj, 457, 355, \dodoi{10.1086/176735}

\bibitem[{{Isella} {et~al.}(2009){Isella}, {Carpenter}, \& {Sargent}}]{isella2009}
{Isella}, A., {Carpenter}, J.~M., \& {Sargent}, A.~I. 2009, \apj, 701, 260, \dodoi{10.1088/0004-637X/701/1/260}

\bibitem[{{Johansen} \& {Youdin}(2007)}]{johansen2007}
{Johansen}, A., \& {Youdin}, A. 2007, \apj, 662, 627, \dodoi{10.1086/516730}

\bibitem[{{Kida}(1981)}]{kida1981}
{Kida}, S. 1981, Journal of the Physical Society of Japan, 50, 3517, \dodoi{10.1143/JPSJ.50.3517}

\bibitem[{{Klahr} {et~al.}(2023){Klahr}, {Baehr}, \& {Melon Fuksman}}]{klahr2023}
{Klahr}, H., {Baehr}, H., \& {Melon Fuksman}, J.~D. 2023, arXiv e-prints, arXiv:2305.08165, \dodoi{10.48550/arXiv.2305.08165}

\bibitem[{{Klahr} \& {Hubbard}(2014)}]{klahr2014}
{Klahr}, H., \& {Hubbard}, A. 2014, \apj, 788, 21, \dodoi{10.1088/0004-637X/788/1/21}

\bibitem[{{Latter}(2016)}]{latter2016}
{Latter}, H.~N. 2016, \mnras, 455, 2608, \dodoi{10.1093/mnras/stv2449}

\bibitem[{{Latter} {et~al.}(2010){Latter}, {Bonart}, \& {Balbus}}]{latter2010b}
{Latter}, H.~N., {Bonart}, J.~F., \& {Balbus}, S.~A. 2010, \mnras, 405, 1831, \dodoi{10.1111/j.1365-2966.2010.16556.x}

\bibitem[{{Latter} \& {Papaloizou}(2017)}]{lp2017}
{Latter}, H.~N., \& {Papaloizou}, J. 2017, \mnras, 472, 1432, \dodoi{10.1093/mnras/stx2038}

\bibitem[{{Lehmann} \& {Lin}(2022)}]{lehmann2022}
{Lehmann}, M., \& {Lin}, M.~K. 2022, \aap, 658, A156, \dodoi{10.1051/0004-6361/202142378}

\bibitem[{{Lehmann} \& {Lin}(2023)}]{lehmann2023}
{Lehmann}, M., \& {Lin}, M.-K. 2023, \mnras, 522, 5892, \dodoi{10.1093/mnras/stad1349}

\bibitem[{{Lesur} \& {Papaloizou}(2009)}]{lesur2009}
{Lesur}, G., \& {Papaloizou}, J.~C.~B. 2009, \aap, 498, 1, \dodoi{10.1051/0004-6361/200811577}

\bibitem[{{Lesur} \& {Papaloizou}(2010)}]{lesur2010}
{Lesur}, G., \& {Papaloizou}, J.~C.~B. 2010, \aap, 513, A60, \dodoi{10.1051/0004-6361/200913594}

\bibitem[{{Lesur} {et~al.}(2023){Lesur}, {Ercolano}, {Flock}, {Lin}, {Yang}, {Barranco}, {Benitez-Llambay}, {Goodman}, {Johansen}, {Klahr}, {Laibe}, {Lyra}, {Marcus}, {Nelson}, {Squire}, {Simon}, {Turner}, {Umurhan}, \& {Youdin}}]{lesur2023}
{Lesur}, G., {Ercolano}, B., {Flock}, M., {et~al.} 2023, in Protostars and Planets VII, ed. S.-i. {Inutsuka}, T.~{Muto}, K.~{Tomida}, \& M.~{Tamura}, 465, \dodoi{10.48550/arXiv.2203.09821}

\bibitem[{{Li} \& {Youdin}(2021)}]{li2021}
{Li}, R., \& {Youdin}, A.~N. 2021, \apj, 919, 107, \dodoi{10.3847/1538-4357/ac0e9f}

\bibitem[{{Lin}(2021)}]{lin2021}
{Lin}, M.-K. 2021, \apj, 907, 64, \dodoi{10.3847/1538-4357/abcd9b}

\bibitem[{{Lin} \& {Youdin}(2015)}]{lin2015}
{Lin}, M.-K., \& {Youdin}, A.~N. 2015, \apj, 811, 17, \dodoi{10.1088/0004-637X/811/1/17}

\bibitem[{{Long} {et~al.}(2018){Long}, {Pinilla}, {Herczeg}, {Harsono}, {Dipierro}, {Pascucci}, {Hendler}, {Tazzari}, {Ragusa}, {Salyk}, {Edwards}, {Lodato}, {van de Plas}, {Johnstone}, {Liu}, {Boehler}, {Cabrit}, {Manara}, {Menard}, {Mulders}, {Nisini}, {Fischer}, {Rigliaco}, {Banzatti}, {Avenhaus}, \& {Gully-Santiago}}]{long18}
{Long}, F., {Pinilla}, P., {Herczeg}, G.~J., {et~al.} 2018, \apj, 869, 17, \dodoi{10.3847/1538-4357/aae8e1}

\bibitem[{{Lovelace} {et~al.}(1999){Lovelace}, {Li}, {Colgate}, \& {Nelson}}]{lovelace1999}
{Lovelace}, R.~V.~E., {Li}, H., {Colgate}, S.~A., \& {Nelson}, A.~F. 1999, \apj, 513, 805, \dodoi{10.1086/306900}

\bibitem[{{Lubow} \& {Pringle}(1993)}]{lubow1993}
{Lubow}, S.~H., \& {Pringle}, J.~E. 1993, \apj, 409, 360, \dodoi{10.1086/172669}

\bibitem[{{Lynden-Bell} \& {Pringle}(1974)}]{lynden1974}
{Lynden-Bell}, D., \& {Pringle}, J.~E. 1974, \mnras, 168, 603, \dodoi{10.1093/mnras/168.3.603}

\bibitem[{{Lyra}(2014)}]{lyra2014}
{Lyra}, W. 2014, \apj, 789, 77, \dodoi{10.1088/0004-637X/789/1/77}

\bibitem[{{Lyra} \& {Klahr}(2011)}]{lyra2011}
{Lyra}, W., \& {Klahr}, H. 2011, \aap, 527, A138, \dodoi{10.1051/0004-6361/201015568}

\bibitem[{{Lyra} \& {Lin}(2013)}]{lyra2013}
{Lyra}, W., \& {Lin}, M.-K. 2013, \apj, 775, 17, \dodoi{10.1088/0004-637X/775/1/17}

\bibitem[{{Lyra} {et~al.}(2018){Lyra}, {Raettig}, \& {Klahr}}]{lyra2018}
{Lyra}, W., {Raettig}, N., \& {Klahr}, H. 2018, Research Notes of the American Astronomical Society, 2, 195, \dodoi{10.3847/2515-5172/aaeac9}

\bibitem[{{Malygin} {et~al.}(2017){Malygin}, {Klahr}, {Semenov}, {Henning}, \& {Dullemond}}]{malygin2017}
{Malygin}, M.~G., {Klahr}, H., {Semenov}, D., {Henning}, T., \& {Dullemond}, C.~P. 2017, \aap, 605, A30, \dodoi{10.1051/0004-6361/201629933}

\bibitem[{{Manger} \& {Klahr}(2018)}]{manger2018}
{Manger}, N., \& {Klahr}, H. 2018, \mnras, 480, 2125, \dodoi{10.1093/mnras/sty1909}

\bibitem[{{Manger} {et~al.}(2020){Manger}, {Klahr}, {Kley}, \& {Flock}}]{manger2020}
{Manger}, N., {Klahr}, H., {Kley}, W., \& {Flock}, M. 2020, \mnras, 499, 1841, \dodoi{10.1093/mnras/staa2943}

\bibitem[{{Mathews} {et~al.}(2013){Mathews}, {Klaassen}, {Juh{\'a}sz}, {Harsono}, {Chapillon}, {van Dishoeck}, {Espada}, {de Gregorio-Monsalvo}, {Hales}, {Hogerheijde}, {Mottram}, {Rawlings}, {Takahashi}, \& {Testi}}]{mathews2013}
{Mathews}, G.~S., {Klaassen}, P.~D., {Juh{\'a}sz}, A., {et~al.} 2013, \aap, 557, A132, \dodoi{10.1051/0004-6361/201321600}

\bibitem[{{Nelson} {et~al.}(2013){Nelson}, {Gressel}, \& {Umurhan}}]{nelson2013}
{Nelson}, R.~P., {Gressel}, O., \& {Umurhan}, O.~M. 2013, \mnras, 435, 2610, \dodoi{10.1093/mnras/stt1475}

\bibitem[{{Paardekooper} {et~al.}(2010){Paardekooper}, {Lesur}, \& {Papaloizou}}]{paardekooper2010}
{Paardekooper}, S.-J., {Lesur}, G., \& {Papaloizou}, J. C.~B. 2010, \apj, 725, 146, \dodoi{10.1088/0004-637X/725/1/146}

\bibitem[{{Petersen} {et~al.}(2007{\natexlab{a}}){Petersen}, {Julien}, \& {Stewart}}]{peterson07a}
{Petersen}, M.~R., {Julien}, K., \& {Stewart}, G.~R. 2007{\natexlab{a}}, \apj, 658, 1236, \dodoi{10.1086/511513}

\bibitem[{{Petersen} {et~al.}(2007{\natexlab{b}}){Petersen}, {Stewart}, \& {Julien}}]{peterson07b}
{Petersen}, M.~R., {Stewart}, G.~R., \& {Julien}, K. 2007{\natexlab{b}}, \apj, 658, 1252, \dodoi{10.1086/511523}

\bibitem[{{Pfeil} \& {Klahr}(2019)}]{pfeil2019}
{Pfeil}, T., \& {Klahr}, H. 2019, \apj, 871, 150, \dodoi{10.3847/1538-4357/aaf962}

\bibitem[{{Raettig} {et~al.}(2015){Raettig}, {Klahr}, \& {Lyra}}]{raettig2015}
{Raettig}, N., {Klahr}, H., \& {Lyra}, W. 2015, \apj, 804, 35, \dodoi{10.1088/0004-637X/804/1/35}

\bibitem[{{Raettig} {et~al.}(2013){Raettig}, {Lyra}, \& {Klahr}}]{raettig2013}
{Raettig}, N., {Lyra}, W., \& {Klahr}, H. 2013, \apj, 765, 115, \dodoi{10.1088/0004-637X/765/2/115}

\bibitem[{{Raettig} {et~al.}(2021){Raettig}, {Lyra}, \& {Klahr}}]{raettig2021}
{Raettig}, N., {Lyra}, W., \& {Klahr}, H. 2021, \apj, 913, 92, \dodoi{10.3847/1538-4357/abf739}

\bibitem[{{Safronov}(1972)}]{safronov1972}
{Safronov}, V.~S. 1972, {Evolution of the protoplanetary cloud and formation of the earth and planets.}

\bibitem[{{Shu}(1984)}]{shu1984}
{Shu}, F.~H. 1984, in Planetary Rings, 513--561

\bibitem[{{Simon} {et~al.}(2016){Simon}, {Armitage}, {Li}, \& {Youdin}}]{simon2016}
{Simon}, J.~B., {Armitage}, P.~J., {Li}, R., \& {Youdin}, A.~N. 2016, \apj, 822, 55, \dodoi{10.3847/0004-637X/822/1/55}

\bibitem[{{Svanberg} {et~al.}(2022){Svanberg}, {Cui}, \& {Latter}}]{svanberg2022}
{Svanberg}, E., {Cui}, C., \& {Latter}, H.~N. 2022, \mnras, 514, 4581, \dodoi{10.1093/mnras/stac1598}

\bibitem[{{Tazzari} {et~al.}(2017){Tazzari}, {Testi}, {Natta}, {Ansdell}, {Carpenter}, {Guidi}, {Hogerheijde}, {Manara}, {Miotello}, {van der Marel}, {van Dishoeck}, \& {Williams}}]{tazzari2017}
{Tazzari}, M., {Testi}, L., {Natta}, A., {et~al.} 2017, \aap, 606, A88, \dodoi{10.1051/0004-6361/201730890}

\bibitem[{{Teed} \& {Latter}(2021)}]{teed2021}
{Teed}, R.~J., \& {Latter}, H.~N. 2021, \mnras, \dodoi{10.1093/mnras/stab2311}

\bibitem[{{Torrence} \& {Compo}(1998)}]{torrence1998}
{Torrence}, C., \& {Compo}, G.~P. 1998, Bulletin of the American Meteorological Society, 79, 61

\bibitem[{{Turner} \& {Drake}(2009)}]{turner2009}
{Turner}, N.~J., \& {Drake}, J.~F. 2009, \apj, 703, 2152, \dodoi{10.1088/0004-637X/703/2/2152}

\bibitem[{{Turner} {et~al.}(2014){Turner}, {Fromang}, {Gammie}, {Klahr}, {Lesur}, {Wardle}, \& {Bai}}]{turner2014}
{Turner}, N.~J., {Fromang}, S., {Gammie}, C., {et~al.} 2014, in Protostars and Planets VI, ed. H.~{Beuther}, R.~S. {Klessen}, C.~P. {Dullemond}, \& T.~{Henning}, 411, \dodoi{10.2458/azu\_uapress\_9780816531240-ch018}

\bibitem[{{Umurhan} {et~al.}(2020){Umurhan}, {Estrada}, \& {Cuzzi}}]{umurhan2020}
{Umurhan}, O.~M., {Estrada}, P.~R., \& {Cuzzi}, J.~N. 2020, \apj, 895, 4, \dodoi{10.3847/1538-4357/ab899d}

\bibitem[{{Urpin}(2003)}]{urpin2003}
{Urpin}, V. 2003, \aap, 404, 397, \dodoi{10.1051/0004-6361:20030513}

\bibitem[{{Urpin} \& {Brandenburg}(1998)}]{urpin1998}
{Urpin}, V., \& {Brandenburg}, A. 1998, \mnras, 294, 399, \dodoi{10.1046/j.1365-8711.1998.01118.x}

\bibitem[{{van der Marel} {et~al.}(2021){van der Marel}, {Birnstiel}, {Garufi}, {Ragusa}, {Christiaens}, {Price}, {Sallum}, {Muley}, {Francis}, \& {Dong}}]{vdmarel2021}
{van der Marel}, N., {Birnstiel}, T., {Garufi}, A., {et~al.} 2021, \aj, 161, 33, \dodoi{10.3847/1538-3881/abc3ba}

\bibitem[{{Youdin} \& {Johansen}(2007)}]{youdin2007}
{Youdin}, A., \& {Johansen}, A. 2007, \apj, 662, 613, \dodoi{10.1086/516729}

\bibitem[{{Youdin} \& {Goodman}(2005)}]{youdin2005}
{Youdin}, A.~N., \& {Goodman}, J. 2005, \apj, 620, 459, \dodoi{10.1086/426895}

\end{thebibliography}
\bibliographystyle{aasjournal}

\end{document}